\documentclass[10pt]{article}

\usepackage{amsmath,amssymb,amsbsy,amsthm,color,epsfig,mathrsfs}
\usepackage{amsbsy,floatrow}
\usepackage{caption}
\usepackage{pifont}

\usepackage{hyperref}
\usepackage{bookmark}

\newtheorem{theorem}{Theorem}

\newtheorem{corollary}[theorem]{Corollary}

\newcounter{spslist}

\newcommand{\mat}[5]{ \renewcommand{\arraystretch}{#1}
                    \left[\! \begin{array}{cc}
                            #2 & #3 \\
                            #4 & #5 \end{array} \!\right] }

\newcounter{geqncount}
    {\refstepcounter{equation}%
     \setcounter{geqncount}{\value{equation}}%
     \setcounter{equation}{0}%
  }%
    {\setcounter{equation}{\value{geqncount}}}

\newcommand{\bigxor}{\mathop{\mathchoice
  {\textstyle\bigoplus}{\textstyle\bigoplus}
  {\scriptstyle\bigoplus}{\scriptscriptstyle\bigoplus}}}

\newcommand{\tensor}{\!\otimes\!}
\newcommand{\C}{{K}}
\newcommand{\Centry}{{\kappa}}
\newcommand{\HH}{{\mathcal H}}
\renewcommand{\AA}{{\mathcal A}}
\newcommand{\CC}{\mathbb{C}}
\newcommand{\ZZ}{\mathbb{Z}}
\newcommand{\RR}{\mathbb{R}}
\newcommand{\TT}{\mathbb{T}}

\newcommand{\slantfrac}[2]{{\hbox{\small$\raisebox{1pt}{#1\!/}\raisebox{-1pt}{\!#2}$}}}
\newcommand{\onehalf}{\slantfrac{1}{2}}

\begin{document}

\bibliographystyle{plain} 

\begin{center}
{\bf \Large  Eigenfunctions of Unbounded Support for Embedded Eigenvalues \\\vspace{1ex} of Locally Perturbed Periodic Graph Operators}
\end{center}

\vspace{0.2ex}

\begin{center}
{\scshape \large Stephen P. Shipman} \\
\vspace{1ex}
{\itshape Department of Mathematics, Louisiana State University\\
Baton Rouge, LA 70803, USA}
\end{center}

\vspace{3ex}
\centerline{\parbox{0.9\textwidth}{
{\bf Abstract.}\
It is known that, if a locally perturbed periodic self-adjoint operator on a combinatorial or quantum graph admits an eigenvalue embedded in the continuous spectrum, then the associated eigenfunction is compactly supported---that is, if the Fermi surface is irreducible, which occurs generically in dimension two or higher.  This article constructs a class of operators whose Fermi surface is reducible for all energies by coupling several periodic systems.  The components of the Fermi surface correspond to decoupled spaces of hybrid states, and in certain frequency bands,
some components contribute oscillatory hybrid states (corresponding to spectrum) and other components contribute only exponential ones.
This separation allows a localized defect to suppress the oscillatory (radiation) modes and retain the evanescent ones, thereby leading to embedded eigenvalues whose associated eigenfunctions decay exponentially but are not compactly supported.
}}

\vspace{3ex}
\noindent
\begin{mbox}
\end{mbox}
\vspace{3ex}

\hrule
\vspace{4ex}

If a periodic self-adjoint difference or differential operator $A$ on a combinatorial or quantum graph is perturbed by a localized operator $V$, and if $A+V$ admits an eigenvalue embedded in the continuous spectrum, then the corresponding eigenfunction (bound state) typically has compact support~\cite{KuchmentVainberg2006}.  
The obstruction to unbounded support is the algebraic fact that a generic polynomial in several variables cannot be factored.  This is reflected in the irreducibility of the Floquet (Fermi) surface of $A$, which is the zero set of a Laurent polynomial $\Delta_\lambda(z)\!=\!0$ that describes the complex vectors $z$ for which $Au=\lambda u$ admits a quasi-periodic solution $u$ with quasi-momentum vector $(k_1,\dots,k_n)\in\CC^n$, where $z=(e^{ik_1},\dots,e^{ik_n})\in{\CC^*}^n$ is the vector of Floquet multipliers.

The Fermi surface is known to be irreducible for all but finitely many energies $\lambda$ for the discrete 2D Laplacian plus a periodic potential~\cite{GiesekerKnorrerTrubowitz1993} and for the continuous Laplacian plus a potential that is separable in a specific way in 2D and 3D~\cite{BattigKnorrerTrubowitz1991,KuchmentVainberg2000}.  In the latter case, the principle of unique continuation of solutions of elliptic equations precludes the emergence of eigenfunctions of compact support under local perturbations.  Thus no embedded eigenvalues are possible.  But unique continuation fails for periodic combinatorial graph operators and quantum graphs~\cite{BerkolaikoKuchment2013,Kuchment2005a} and for higher-order elliptic equations~\cite{Kuchment1993}.  In these cases, spectrally embedded eigenfunctions with compact support do exist, even for unperturbed periodic operators.
In the graph case, they can be created by attaching a finite graph to the periodic one at a vertex of the finite graph where one of its eigenfunctions vanishes.

This article constructs a class of periodic graph operators for which the Fermi surface is {\em reducible} for all energies and for which local perturbations create embedded eigenvalues whose eigenfunctions have {\em unbounded} support.  These operators are constructed by coupling $m$ different operators on identical graphs.  The resulting operator decouples into $m$ invariant subspaces of hybrid states with different spectral bands.  A non-embedded eigenvalue for one of these hybrid spaces that lies in a spectral band of another is an embedded eigenvalue for the full system.  A simple example is two copies of the integer lattice $\ZZ^2$, placed one atop the other, endowed with the discrete Laplace operator, or the quantum version in which edges connect adjacent vertices.  The reducibility of the Fermi surface for all energies is automatic: each of its components corresponds to an invariant subspace of the operator.

Questions on the analytic structure of the Fermi surface, in particular the determination of (ir)reducibility, are not easy
(see~\cite{KnorrerTrubowitz1990}, for example).  Reducibility for the class of operators in the present work results intentionally from its explicit construction.
Each irreducible component is contained in the Fermi surface for an invariant subspace of the graph operator.  If a component corresponding to an invariant subspace fails to intersect $\RR^n$ at an energy $\lambda$, then $\lambda$ is not in the spectrum for that subspace and one can create a defect that supports an eigenfunction (bound state) within that subspace.  This evokes the question of whether each irreducible component of the Fermi surface always corresponds to an invariant subspace of the operator, because this would raise the prospect of creating a defect that produces an eigenvalue whenever at least one irreducible component of the Fermi surface does not intersect $\RR^n$.  This was conjectured for Schr\"odinger operators in \cite[\S5,\,point\,3]{KuchmentVainberg2000}.

The Fermi ``surface" of a 1-periodic combinatorial or quantum graph operator or ODE is always reducible; its components are simply the roots $z_j$ of the Laurent polynomial $\Delta_\lambda(z)$ of one variable $z\in\CC$.
It is easy to construct embedded eigenvalues with exponentially decaying eigenfunctions because of the explicit decoupling of the Floquet modes $u_jz_j^{g}$, where $g\in\ZZ$ and $u_j$ is the restriction of the mode to one period ({\itshape e.g.} \cite{AyaCanoZhevandrov2012,ShipmanRibbeckSmith2010,ShipmanWelters2013,Tillay2012})---one splices an exponentially growing mode to the left of a defect together with an exponentially decaying mode to the right.
An examination of some 1D examples that can be computed by hand motivates the constructions in higher dimensions.

\section{Embedded eigenvalues in 1-periodic graphs} 

The purpose of this section is to illustrate the ideas of the paper through three examples of 1D graph operators for which one can straightforwardly compute spectrally embedded bound states of unbounded support.
{\bfseries Example~1} shows how embedded eigenvalues are easily created in 1D periodic graphs simply because the Laurent polynomial $\Delta_\lambda(z)$ is generically a product of linear factors, where $\Delta_\lambda(e^{ik})=0$ is the dispersion relation between energy $\lambda$ and quasi-momentum (or wavenumber) $k$.  The construction does not generalize to higher dimensions, where $\Delta_\lambda(z)$ generically fails to factor.
{\bfseries Example~2} for a combinatorial graph does generalize to higher dimensions (sec.~\ref{sec:combinatorial}) because the construction of bound states is devised specifically to be independent of dimension.  It relies on an explicit decoupling of a graph operator into two independent subsystems with different continuous spectrum.
{\bfseries Example~3} shows how to modify Example~2 to accommodate quantum graphs; it is generalized to higher dimensions in section~\ref{sec:quantum}.

\subsection{Example 1: Finite-difference operator of order 4}

\begin{figure}  
\floatbox[{\capbeside\thisfloatsetup{capbesideposition={right,top},capbesidewidth=6.1cm}}]{figure}[\FBwidth]
{\caption{\small This is the dispersion relation $\lambda=\hat A(e^{ik})=4\cos k + 2\cos 2k$ for the fourth-order difference operator $A$ in Example~1.  It characterizes solutions of the form $u(g)=e^{ikg}$ to the equation $Au=\lambda u$.  The spectrum $\sigma(A)$ of $A$ is the range $[-3,6]$ of this graph; it has multiplicity 4 in $(-3,-2)$ and multiplicity 2 in $(-2,6)$.}
\label{fig:Dispersion1}}
{\scalebox{0.42}{\includegraphics{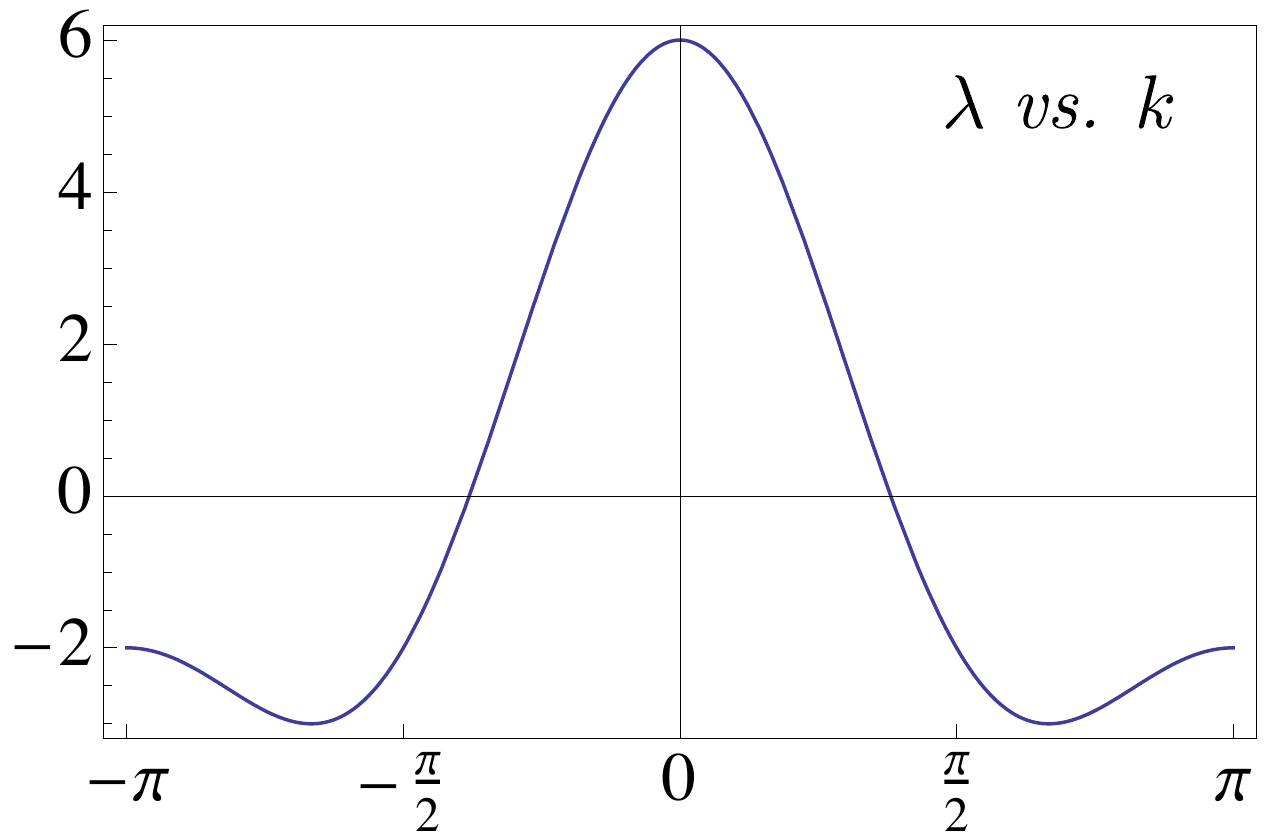}}\hspace{2em}}
\end{figure}

Consider the fourth-order difference operator $A$ on $\ell^2(\ZZ)$ given by
\begin{equation*}
  (Au)(g) \,=\, 2\big(u(g+1)+u(g-1)\big) + \big(u(g+2)+u(g-2)\big),
  \qquad g\in\ZZ.
\end{equation*}
The $z$-transform $u\mapsto\hat u$ ({\itshape i.e.}, the Floquet transform $\hat u(g,z)$ evaluated at $g=0$),
\begin{equation*}
  \hat u(z) = \sum_{g\in\ZZ} u(g) z^{-g}\,,
\end{equation*}
converts $A$ into a multiplication operator
\begin{eqnarray*}
  && (Au)\hat{}\,(z) \,=\, \hat A(z) \hat u(z) \\
  && \hat A(z) \,=\, 2(z+z^{-1}) + (z^2+z^{-2})\,.
\end{eqnarray*}
It is a Hilbert-space isomorphism from $\ell^2(\ZZ)$ to $L^2(\TT)$, where $\TT$ is the complex unit circle $\TT=\{z\in\CC\,:\,|z|=1\}$.
This shows that the spectrum $\sigma(A)$ of $A$ consists of those $\lambda$ for which $\hat A(e^{ik})=\lambda$ for some $k\in\RR$.  This ``dispersion relation" between $\lambda$ and $k$,
\begin{equation*}
  \lambda = \hat A(e^{ik}) = 4\cos k + 2\cos 2k
\end{equation*}
is shown in Fig.~\ref{fig:Dispersion1} for real $k$.  The spectrum of $A$ is the range $[-3,6]$ of this trigonometric polynomial in $k$.

As seen in Fig.~\ref{fig:Dispersion1}, for $-2<\lambda<6$ the spectrum is of multiplicity 2---there is exactly one pair of solutions of $\hat A(z)=\lambda$ of the form $z=e^{\pm ik}$ with $0<k<\pi$.  This can be seen algebraically by writing $\hat A(z)=\lambda$ as
\begin{equation}\label{zz}
  z+z^{-1} \,=\, -1\pm\sqrt{3+\lambda\,}\,.
\end{equation}
Each choice of sign of the square root gives a pair of solutions of the form $z^{\pm1}$, which are of unit modulus if and only if $|\!-\!1\pm\sqrt{3+\lambda\,}|\leq2$.  In the $\lambda$-interval $(-2,6)$, the plus sign yields $z=e^{\pm ik}$ ($0<k<\pi$) and the minus sign yields $z=-e^{\pm\alpha}$, with $\alpha>0$.

This means that there are both oscillatory solutions $u(g)\!=\!e^{\pm ikg}$ and exponential solutions $u(g)\!=\!(-e^{\pm\alpha})^g$ of $(A-\lambda I)u=0$.  This is because $A$ acts on fields of the form $\chi_z(g)=z^g$ (eigenfunctions of the shift operator not in $\ell^2(\ZZ)$) by multiplication by $\hat A(z)$:
\begin{equation*}
  (A\chi_z)(g) = \hat A(z) \chi_z(g)\,.
\end{equation*}
In this spectral interval, $-2<\lambda<6$, the exponential solutions can be used to construct a spectrally embedded eigenfunction (bound state) $v$ for a localized perturbation $A+V$ of $A$, by splicing an exponentially growing solution for $g\leq0$ with an exponentially decaying one for $g\geq0$,
\begin{equation*}
  v(g) := \renewcommand{\arraystretch}{1.3}
\left\{
  \begin{array}{ll}
    (-e^\alpha)^g\, & g\leq0\,, \\
    (-e^\alpha)^{-g}\, & g\geq0\,.
  \end{array}
\right.
\end{equation*}
Let the potential $V$ be given by a multiplication operator
\begin{equation*}
  (Vu)(g) = V_g\, u(g)
\end{equation*}
with $V_g=0$ for all but finitely many values of $g$.
By enforcing the equation $(A+V)u =\lambda u$, one obtains $V_g=0$ for $|g|\geq2$ and
\begin{equation}
  \renewcommand{\arraystretch}{1.1}
\left.
  \begin{array}{rl}
    V_0 \;= & \lambda + 4e^{-\alpha} - 2e^{-2\alpha} \\
    V_{-1} = V_{1} \;= & \lambda - 2(2-e^{-\alpha})\cosh\alpha\,.
  \end{array}
\right.
\end{equation}
A typical perturbation of $V$ will destroy the bound state and the embedded eigenvalue of $A+V$, resulting in resonant scattering of the extended eigenstates $e^{\pm ikg}$~\cite{Tillay2012}.


\subsection{Example 2: Decoupling by symmetry in a combinatorial graph}
The construction of Example~1 does not extend to $n$-periodic graphs for $n>1$ because the {\em Floquet surface}\footnote{The complex dispersion relation between energy $\lambda$ and quasi-momentum $k$ is $\Delta_\lambda(e^{ik}):=\det(\hat A(e^{ik})-\lambda)=0$, and this zero-set of $(k,\lambda)$-values is the Bloch variety; the Fermi surface for an energy $\lambda$ is $\{k : \Delta_\lambda(e^{ik})=0)\}$, and the Floquet surface for $\lambda$ is $\{z : \Delta_\lambda(z)=0)\}$ \cite{KuchmentVainberg2006}.}
for $\lambda$, $\{z:\hat A(z)-\lambda=0\}$ (more generally, $\{z:\det(\hat A(z)-\lambda)=0\}$) is generically irreducible over $z\in\CC^n$.
Example~2 illustrates a construction, which generalizes to a class of $n$-periodic combinatorial graph operators (sec.~\ref{sec:combinatorial}), for which the Floquet surface is reducible for all $\lambda$ and embedded eigenvalues with eigenfunctions of unbounded support can be created by local defects.

\begin{figure}  
\floatbox[{\capbeside\thisfloatsetup{capbesideposition={right,top},capbesidewidth=6.5cm}}]{figure}[\FBwidth]
{\caption{\small The combinatorial graph $\Gamma$ of Example~2.  The edges indicate interactions between neighboring vertices realized by the operator $A$ (\ref{A2}) that acts on $\ell^2(\mathrm{vert}(\Gamma))$.}
\label{fig:Graph1}}
{\scalebox{0.3}{\includegraphics{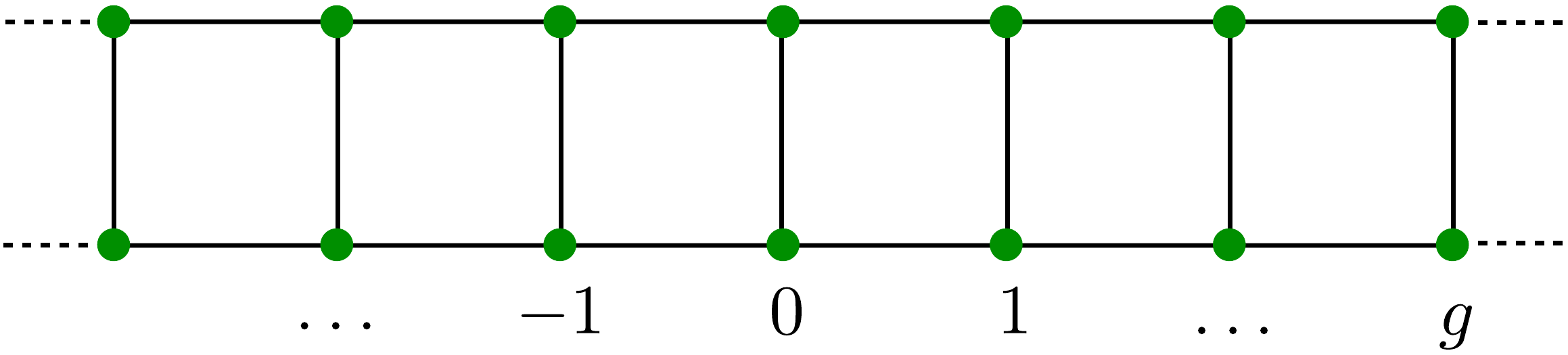}}\hspace{1em}}
\end{figure}

The combinatorial graph $\Gamma$ in Fig.~\ref{fig:Graph1} consists of two coupled 1D chains.
A function $u$ on the vertex set of $\Gamma$ can be viewed as a $\CC^2$-valued function of $g\in\ZZ$.
The edges of $\Gamma$ indicate interactions between neighboring vertices realized by a periodic self-adjoint operator $A$ on $\ell^2(\mathrm{vert}(\Gamma))$:
\begin{eqnarray}
  && (Au)(g) \,=\,  A_0 u(g) + u(g+1) + u(g-1)\,, \label{A2}\\
  && A_0 = \mat{1.1}{a+b}{c}{c}{a-b}\,, \qquad (a,b,c\,\in\RR)\,.
\end{eqnarray}
Under the $z$-transform, $A$ becomes multiplication by a matrix function $\hat A(z)$,
\begin{eqnarray}
  && (Au)\hat{}\,(z) \,=\, \hat A(z) \hat u(z) \\
  && \hat A(z) \,=\, (z+z^{-1}) I + A_0 \,=\, \mat{1.3}{a+(z+z^{-1})+b}{c}{c}{a+(z+z^{-1})-b}\,.
\end{eqnarray}

The equation $(A-\lambda I)u=0$ has a solution of the form\footnote{A function $u(g)=u(0)z^g$ is a non-$L^2$ eigenfunction of the shift operator $u\mapsto u(\cdot+1)$ with eigenvalue $z=e^{ik}$.  The function $u(g)$ is a Floquet-Bloch, or quasi-periodic, solution of $Au=\lambda u$, $z$ is the Floquet multiplier, and $k$ is the quasi-momentum.}
 $u(g)=u(0)z^g$ if and only if $u(0)$ is a null vector of $\hat A(z)-\lambda I$, and the spectrum of $A$ is all $\lambda$ such that $\det(\hat A(e^{ik})-\lambda)=0$ holds for some $k\in\RR$.  The Floquet surface $\det(\hat A(z)-\lambda)=0$ reduces to
\begin{equation*}
  z+z^{-1} \,=\, \lambda - a \pm\sqrt{b^2+c^2\,}\,.
\end{equation*}
By putting $z=e^{ik}$, one obtains two branches of a dispersion relation between energy $\lambda$ and wavenumber $k$.  The parts of these branches where $k$ is real correspond to two $\lambda$-intervals, whose union is the spectrum of $A$,
\begin{eqnarray}
  && \lambda \in (a-2,a+2)+\sqrt{b^2+c^2\,}\,, \notag\\
  && \lambda \in (a-2,a+2)-\sqrt{b^2+c^2\,}\,.\notag
\end{eqnarray}
The quantity $\sqrt{b^2+c^2\,}$ is the magnitude of the splitting of these two {\em energy bands} and is akin to the Rabi frequency.
When the bias $b$ vanishes, the plus-branch has eigenvector $u(0)=[1,1]^t$, corresponding to symmetric solutions $u(g)=u(0)z^{\pm g}$ of $(A-\lambda I)u=0$, and  the minus-branch has eigenvector $u(0)=[1,-1]^t$, corresponding to anti-symmetric solutions.

In the $\lambda$-intervals of multiplicity 2, where the two bands do not overlap, one can create exponentially decaying eigenfunctions for embedded eigenvalues of a locally perturbed operator $A+V$ similarly to Example~1; this is carried out in~\cite{ShipmanRibbeckSmith2010}.  Generic perturbations of $A+V$ destroy the embedded eigenvalue.  In~\cite{ShipmanRibbeckSmith2010}, it is shown that the resulting scattering resonance is detuned from the bound-state energy because of asymmetry created by the bias $b$.

\subsection{Example 3: Decoupling by symmetry in a quantum graph}\,

\begin{figure}  
\scalebox{0.33}{\includegraphics{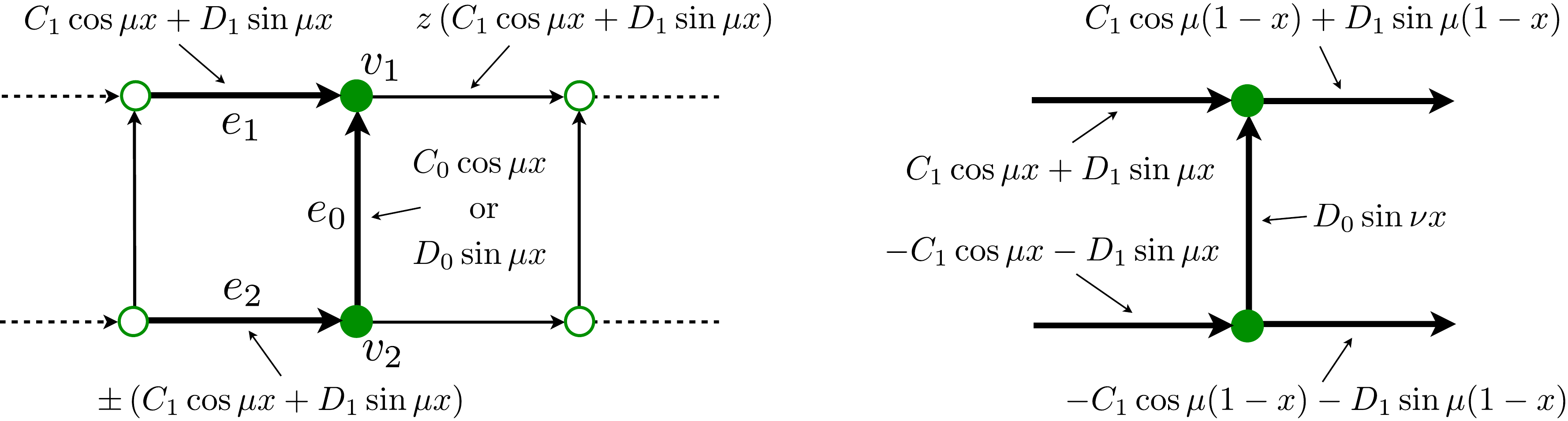}}
\caption{\small {\bfseries Left.} The metric graph $\Gamma$ of Example~3.  A fundamental domain $W$ for its translational symmetry group $\ZZ$ consists of the bold edges $e_1$, $e_2$, $e_0$ and the vertices $v_1$, $v_2$.  Edges $e_1$ and $e_2$ are identified with the $x$-interval $[0,1]$, and edge $e_0$ with $[-1/2,1/2]$ in the direction of the arrows, and the vertices carry no mass.  The functions on the edges are Floquet solutions $u(x)$ of $-d^2u/dx^2 = \lambda u$ ($\lambda=\mu^2$) with Floquet multiplier $z$.  Symmetric and anti-symmetric solutions with respect to a horizontal line are depicted; the $+$ ($-$) sign for $e_2$ corresponds to $C_0\cos\mu x$ ($D_0\cos\mu x$) on $e_0$.
{\bfseries Right.} One vertical edge is made defective by adding a potential so that $-d^2u/dx^2 +V_0 u= \lambda u$ and an antisymmetric solution is $D_0\sin\nu x$ with $\nu=\sqrt{\mu^2-V_0}$.  For spectral values $\lambda=\mu^2$ of multiplicity $2$ around a multiple of $2\pi$ (see Fig.~\ref{fig:Dispersion2}) and appropriate choice of $V_0$, one can construct an $L^2$-eigenfunction that is symmetric about the defective edge and anti-symmetric about the central horizontal line.}
\label{fig:QuantumGraph1}
\end{figure}

Figure~\ref{fig:QuantumGraph1} depicts a simple 1-periodic metric graph $\Gamma$ ({\itshape e.g.} \cite[\S1.3]{BerkolaikoKuchment2013}).
The group $\ZZ$ acts as a translational symmetry group with a fundamental domain $W$ consisting of two vertices $v_1$ and $v_2$, two horizontal edges $e_1$ and $e_2$ coordinatized by $x\in[0,1]$, and a vertical edge $e_0$ coordinatized by $x\in[-1/2,1/2]$.
Define an operator $A$ by
\begin{equation}
  Au(x) \;=\; -\frac{d^2 u}{dx^2}(x) \quad \text{on each edge}.
\end{equation}
$A$ acts on functions $u:\Gamma\to\CC$, such that the restriction of $u$ to each edge $e$ is in the Sobolev space $H^2(e)$, $u$~is continuous at each vertex, and the sum of the derivatives of $u$ at each vertex $v$ directed away from $v$ must vanish ($0$-flux, or Neumann, condition \cite[p.\,14]{BerkolaikoKuchment2013}).  The additional requirement that $|u|^2$ be integrable over $\Gamma$ makes $A$ a self-adjoint operator in~$L^2(\Gamma)$, thus creating a quantum graph $(\Gamma,A)$.

In analogy to Examples 1 and~2, one seeks solutions of $Au=\lambda u$ (not in $L^2(\Gamma)$) that satisfy the quasi-periodic condition
\begin{equation*}
  u(gp) = u(p) z^g
  \qquad \text{for all points $p\in W$ and $g\in\ZZ$.} 
\end{equation*}
On each edge, $u$ has the form $u(x)=C\cos\mu x + D\sin\mu x$, where $\lambda=\mu^2$, as depicted in Fig.~\ref{fig:QuantumGraph1}.
Observe that $A$ is invariant on the symmetric and anti-symmetric spaces of functions with respect to the horizontal line of reflectional symmetry of $\Gamma$.  By requiring that $u$ be anti-symmetric, it has the form $u(x)=D_0\sin\mu x$ on the vertical edge $e_0$, and the continuity and flux conditions at the vertex $v_1$ impose three homogeneous linear conditions on $D_0$ and the coefficients $C_1$ and $D_1$ for the edge $e_1$:
\begin{equation}\label{quantummatrix}
  \renewcommand{\arraystretch}{1.2}
\left[
  \begin{array}{ccc}
    -\sin\frac{\mu}{2} & z & 0 \\
    -\sin\frac{\mu}{2} & \cos\mu & \sin\mu \\
    -\mu\cos\frac{\mu}{2} & \mu\sin\mu & \mu(z-\cos\mu) 
  \end{array}
\right]
\renewcommand{\arraystretch}{1.2}
\left[
  \begin{array}{c}
    D_0 \\ C_1 \\ D_1
  \end{array}
\right]
\,=\,
\renewcommand{\arraystretch}{1.2}
\left[
  \begin{array}{c}
    0 \\ 0 \\ 0
  \end{array}
\right]\,.
\end{equation}
For symmetric functions $u$, one just changes $[D_0,C_1,D_1]^t$ to $[C_0,C_1,D_1]^t$ and, in the first column of the matrix, $\sin\frac{\mu}{2}$ to $\cos\frac{\mu}{2}$ and $\cos\frac{\mu}{2}$ to $-\sin\frac{\mu}{2}$.  Setting the determinants of these matrices to $0$ yields conditions for nonzero quasi-periodic solutions $u$ to $Au=\lambda u$,
\begin{equation}\label{zequations}
  \renewcommand{\arraystretch}{1.1}
\left.
  \begin{array}{ll}
      z + z^{-1} \,=\, 3\cos\mu-1 & \quad\text{symmetric states} \\
      z + z^{-1} \,=\, 3\cos\mu+1 & \quad\text{anti-symmetric states}. \\
  \end{array}
\right.
\end{equation}

\begin{figure}  
\floatbox[{\capbeside\thisfloatsetup{capbesideposition={right,top},capbesidewidth=7cm}}]{figure}[\FBwidth]
{\caption{\small Branches of the dispersion relation $2\cos k = 3\cos\mu\pm1$ (energy $\lambda=\mu^2$) for the operator $A$ of Example~3 on the metric graph in Fig.~\ref{fig:QuantumGraph1} (one period in $\mu$ is shown).  The upper branch corresponds to fields $u(g)=u(0)e^{ikg}$ that are anti-symmetric about the central horizontal line of $\Gamma$; $k$ is real on the spectral band indicated by the upper solid line, where $|2\cos k|\leq2$.  The lower branch corresponds to symmetric fields and the spectral band indicated by the lower solid lines.}
\label{fig:Dispersion2}}
{\scalebox{0.48}{\includegraphics{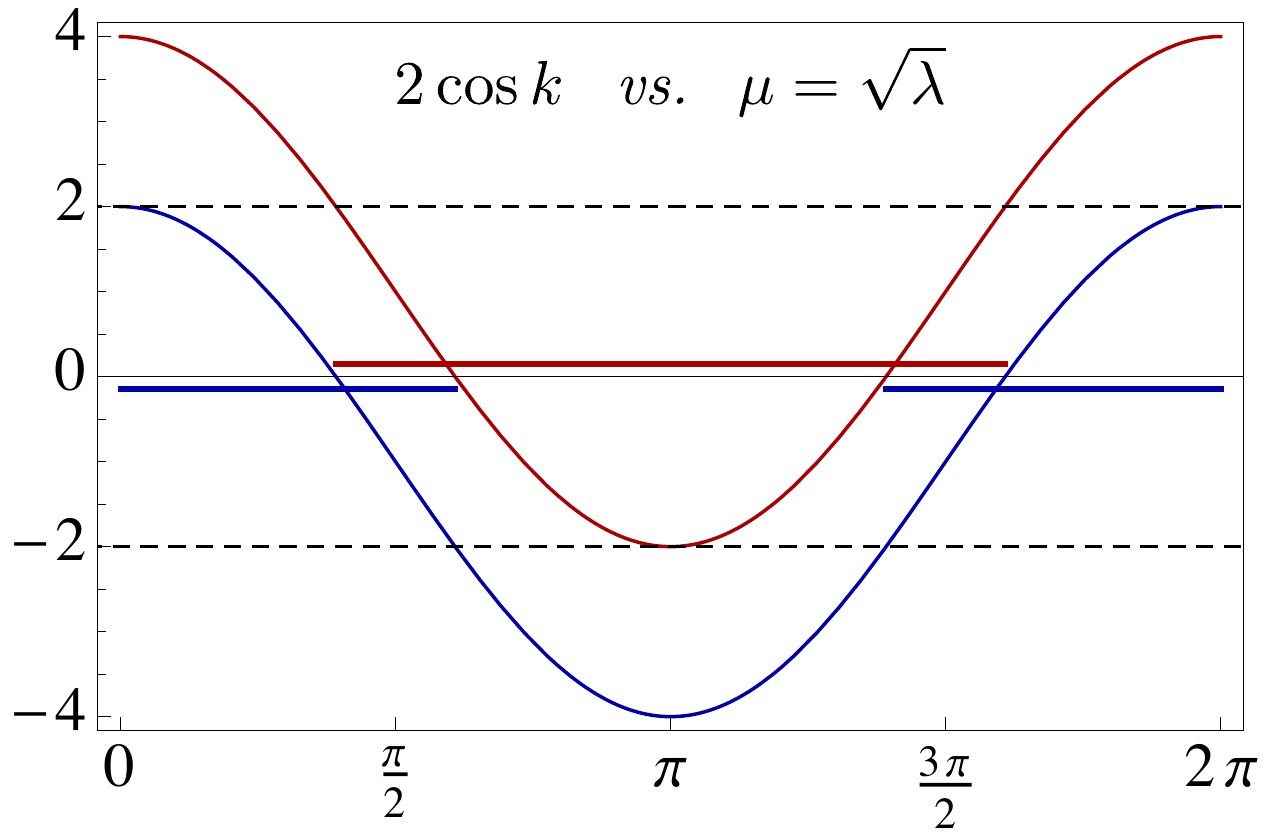}}\hspace{1em}}
\end{figure}

By setting $z=e^{ik}$, one obtains symmetric and antisymmetric branches of the dispersion relation, which are depicted in Fig.~\ref{fig:Dispersion2}.  Each branch exhibits spectral bands, indicated by the solid lines, separated by gaps, but these bands overlap so that the spectrum of $A$ consists of all $\lambda\geq0$.

Let the operator $A$ be perturbed locally by adding to it a potential $V$ that vanishes everywhere except on the edge $e_0$ of the fundamental domain $W$ in Fig.~\ref{fig:QuantumGraph1}.  This means that the action of  $A+V$ is
\begin{eqnarray*}
  && (A+V)u(x) \;=\; -\frac{d^2 u}{dx^2}(x) + V_0u(x) \quad \text{on the $e_0$-edge of $W$,} \\
  && (A+V)u(x) \;=\; -\frac{d^2 u}{dx^2}(x) \hspace{5.2em} \text{on every other edge,}
\end{eqnarray*}
in which $V_0$ is a real number.
Thus solutions of $(A+V)u=\mu^2 u$ have the form $u(x)=C\cos\mu x + D\sin\mu x$ on each edge except $e_0$, where it has the form $u(x)=A\cos\nu x + B\sin\nu x$, where $\nu^2=\mu^2-V_0$.
The continuous spectra of $A$ and $A+V$ are identical because $V$ is a relatively compact perturbation of $A$.

An embedded eigenvalue for the defective quantum graph can be created at spectral values $\lambda=\mu^2$ of multiplicity~2.  This occurs, say, if $\mu$ is near a multiple of $2\pi$ in Fig.~\ref{fig:Dispersion2}, where the symmetric states are propagating ($|z|\!=\!1$) and the anti-symmetric states are exponential ($|z|\!\not=\!1$).  An anti-symmetric eigenfunction (bound state) is created by splicing an exponentially decaying quasi-periodic solution of $Au=\mu^2 u$ to the right of the defective edge with an exponentially growing solution to the left (Fig.~\ref{fig:QuantumGraph1}, right).  Specifically, the second equation of (\ref{zequations})~gives two solutions $z^{\pm1}$, with the smaller one equal to
\begin{equation}\label{zvalue}
  z \;=\; 3\cos\mu + 1 -\sqrt{(3\cos\mu+1)^2-1\,}\,,
  \qquad
  -1<z<1\,.
\end{equation}
The bound state has the form
\begin{equation}\label{quantumboundstate}
  u(gp) =
  \renewcommand{\arraystretch}{1.1}
\left\{
  \begin{array}{ll}
    u_+(p) z^g & \text{for $g>0$} \\
    u_-(p) z^{-g} & \text{for $g\leq0$}
  \end{array}
\right.
  \qquad \text{for all points $p\in W$ and $g\in\ZZ$,} 
\end{equation}
in which $u_+$ and $u_-$ satisfy $Au_\pm = \mu^2 u_\pm$ for $z$ and $z^{-1}$, respectively, subject to~(\ref{zvalue}).
In fact, $u_+(p)z^g$ and $u_-(p)z^{-g}$ are reflections of one another about $e_0$ because of the corresponding reflection symmetry of $\Gamma$ and~$A$.
By setting $u(x)=D_0\sin\nu x$, the continuity and $0$-flux condition at the vertex $v_1$ in Fig.~\ref{fig:QuantumGraph1}
result in a relation between $\nu$ and $\mu$,
\begin{equation}\label{numu}
  \nu\cot{\textstyle\frac{\nu}{2}} \,=\, 2(z-\cos\mu)\frac{\mu}{\sin\mu}\,.
\end{equation}
Remember that $z$ depends on $\mu$ through (\ref{zvalue}) and that $\nu^2=\mu^2-V_0$.  As long as one can solve for $\nu$ in terms of $\mu$ in (\ref{numu}), the potential $V_0$ can be determined so that (\ref{quantumboundstate}) satisfies $(A+V)u=\lambda u$, thus completing the construction of an embedded eigenvalue whose eigenfunction has unbounded support.  This is possible because the left-hand side of~(\ref{numu}) takes on all real values as $\nu$ ranges over~$\RR$.

\section{Embedded eigenvalues in coupled $n$-periodic graphs}\label{sec:combinatorial} 

This section generalizes the 1D Example~2 to higher dimension.
The first step (sec.~\ref{sec:decoupling}) is to {\em couple} two identical combinatorial graphs, with possibly different operators, in such a way that the resulting system {\em decouples} into two spaces of hybrid states with different continuous spectrum.  Next (sec.~\ref{sec:eigenfunctions}), a non-embedded eigenvalue is constructed for one of the hybrid systems with energy in the spectral band of the other.  The construction is generalized to $m$ coupled graphs in section~\ref{sec:several}.

The mathematical development of this coupling-decoupling construction in sections~\ref{sec:decoupling} and~\ref{sec:several} is valid in a general Hilbert-space setting, although it is presented in the language of combinatorial graphs.
In particular, it can be applied to the coupling of two identical quantum graphs.  However, since they are coupled by interactions ``at a distance", the coupled system is not a quantum graph.  Section~\ref{sec:quantum} presents a modification of the construction for quantum graphs.

\subsection{Decoupling of hybrid states in coupled graphs}\label{sec:decoupling}

Let $\Gamma$ be a combinatorial or metric graph that is $n$-periodic, meaning that $\Gamma$ admits a group of symmetries isomorphic to $\ZZ^n$.  Assume also that a fundamental domain $W$ of the $\ZZ^n$ action on $\Gamma$ is pre-compact.  
Let $A$ be a periodic operator on $\Gamma$, whose domain $\mathrm{dom}(A)$ is a dense sub-vector-space of the Hilbert space $\HH$ of square-integrable functions on $\Gamma$, and let $A$ be self-adjoint in $\HH$.  The periodicity of $A$ means that $A$ commutes with the action of $\ZZ^n$.  

Consider two copies of the same graph $\Gamma$, one endowed with the operator $A+B$ and the other with the operator $A-B$, where the {\em bias} $B$ is bounded, periodic, and self-adjoint.  The two systems $(\HH,A+B)$ and $(\HH,A-B)$ are then coupled through a bounded periodic operator $C$ to create a periodic self-adjoint operator $\mathcal A$ on the disjoint union $\Gamma\mathring{\cup}\Gamma$.  The domain of $\mathcal A$ is $\mathrm{dom}(A)\oplus\mathrm{dom}(A)\subset\HH\oplus\HH$, and its block-matrix representation with respect to this decomposition is
\begin{equation}\label{mathcalA}
  \AA \,=\,
  \renewcommand{\arraystretch}{1.3}
\left[
  \begin{array}{cc}
    A+B & C \\
     C^* & A-B
  \end{array}
\right].
\end{equation}

It turns out that, if $B$ and $C$ are linearly dependent operators, then $\mathcal A$ is unitarily block-diagonalizable.
Thus, let $B$ and $C$ be multiples of a given bounded, periodic, self-adjoint operator $L$ on $\mathcal H$:
\begin{equation}\label{biascoupling}
  \renewcommand{\arraystretch}{1.2}
\left.
  \begin{array}{ll}
    B\,=\, \cos(\theta) L & \text{(bias)} \\
    C\,=\, e^{i\phi}\sin(\theta) L & \text{(coupling).}
  \end{array}
\right.
\end{equation}
Here, $\phi$ is an arbitrary phase, and $\theta$ measures the relative strengths of the bias and the coupling.
The operator $\mathcal A$ is decoupled into two operators $A+L$ and $A-L$ by the unitary operator
\begin{equation}\label{U}
  {\mathcal U} \,=\,
  \renewcommand{\arraystretch}{1.5}
\left[
  \begin{array}{cc}
    \cos(\theta/2) I & -e^{i\phi} \sin(\theta/2) I \\
    e^{-i\phi}\sin(\theta/2) I & \cos(\theta/2) I
  \end{array}
\right].
\end{equation}
Indeed, a calculation yields
\begin{eqnarray}\label{AUUA1}
  \AA\,{\mathcal U} \;=\; {\mathcal U}\tilde\AA
\end{eqnarray}
on the domain $\mathrm{dom}(A)\oplus\mathrm{dom}(A)\subset\HH\oplus\HH$, in which
\begin{equation*}
  \tilde\AA \;=\;
  \renewcommand{\arraystretch}{1.5}
\left[
  \begin{array}{cc}
    A+L & 0 \\
    0 & A-L 
  \end{array}
\right].
\end{equation*}
If $A$ is a graph operator, then
\begin{equation*}
  \det(\hat \AA(z)- \lambda) = \det(\hat A(z)+\hat L(z)-\lambda)\det(\hat A(z)-\hat L(z)-\lambda)\,,
\end{equation*}
so that {\em the Floquet surface $\det(\hat \AA(z)- \lambda) =0$ is reducible for all energies $\lambda$.}

\smallskip
The conjugacy~(\ref{AUUA1}) effects a decomposition of $\HH\oplus\HH$ into two orthogonal $\AA$-invariant spaces $\HH_+$ and $\HH_-$ of hybrid states
\begin{eqnarray}
  && \HH_+ \;=\; \left\{ \left( \cos(\textstyle\theta/2)\,u\,,\; e^{-i\phi}\!\sin(\theta/2)\,u \right) \;:\; u\in\HH \right\} \label{Hplus}\\
  && \HH_- \;=\; \left\{ \left( -e^{i\phi}\!\sin(\textstyle\theta/2)\,u\,,\; \cos(\theta/2)\,u \right) \;:\; u\in\HH \right\} \label{Hminus}\\
  && \HH_+\oplus\HH_- \,=\, \HH\oplus\HH\,.\notag
\end{eqnarray}
The action of $\AA$ on $\HH_+$ is given by applying $A+L$ to each component,
\begin{equation}
  \AA\left( \cos(\textstyle\theta/2)\,u\,,\; e^{-i\phi}\!\sin(\theta/2)\,u \right)
  = \left( \cos(\textstyle\theta/2)\,(A+L)u\,,\; e^{-i\phi}\!\sin(\theta/2)\,(A+L)u \right)\,,
\end{equation}
and on $\HH_-$ the action is by $A-L$.

\begin{figure}  
\floatbox[{\capbeside\thisfloatsetup{capbesideposition={right,top},capbesidewidth=7cm}}]{figure}[\FBwidth]
{\caption{\small The relative strengths of the bias and coupling in the operator $\mathcal A$, defined in (\ref{mathcalA}) and (\ref{biascoupling}), is represented by an angle $\theta$.  There are two $\mathcal A$-invariant spaces ${\mathcal H}_+$ and ${\mathcal H}_-$ of hybrid states given by (\ref{Hplus},\ref{Hminus}).  The relative amplitudes of the components $(u_1,u_2)$ of states in ${\mathcal H}_+$ (resp. ${\mathcal H}_-$) are given by the angle $\theta/2$ (resp. $\theta/2-\pi/2$), as indicated by the dots.}
\label{fig:theta}}
{
\raisebox{4ex}{\scalebox{0.4}{\includegraphics{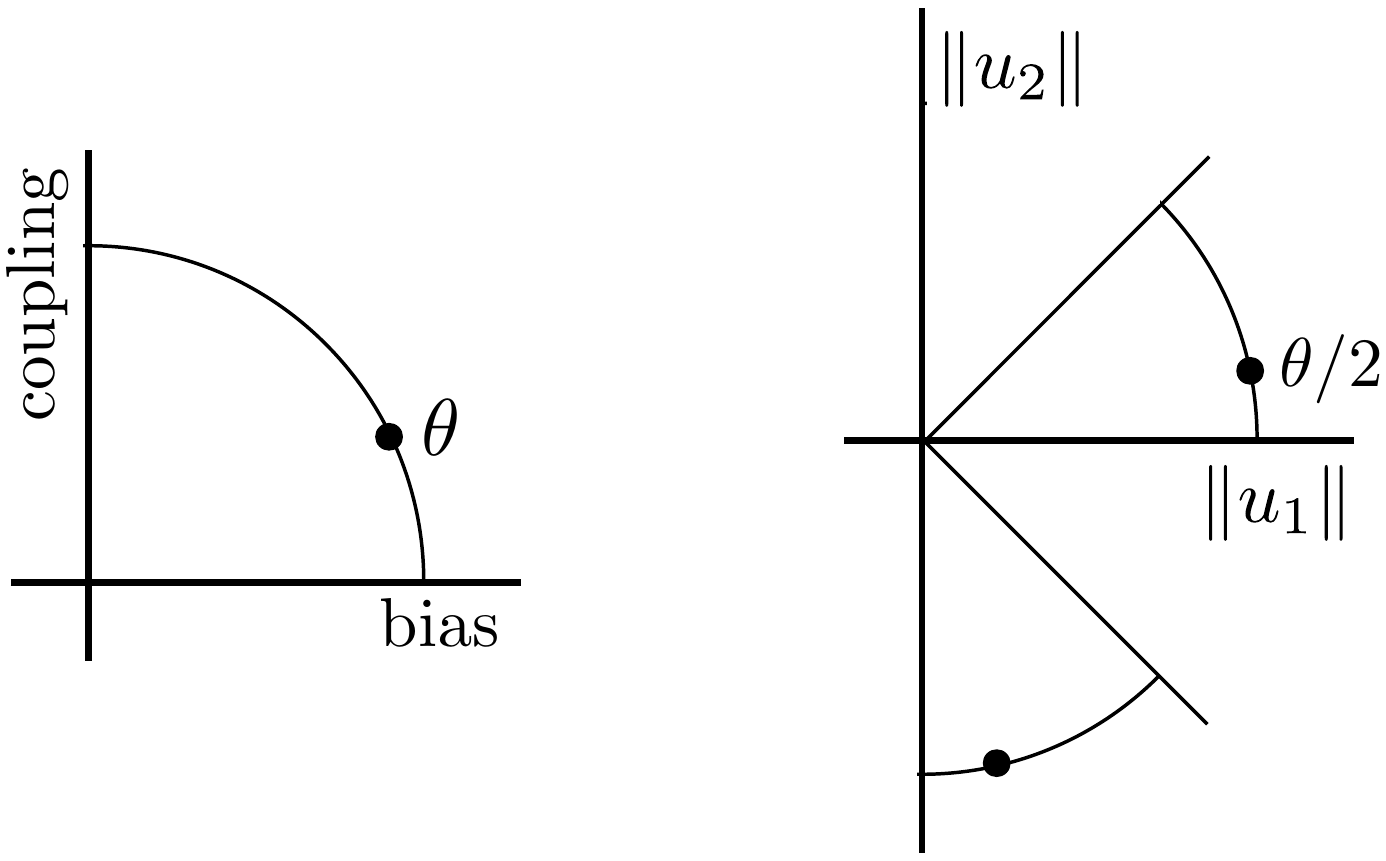}}
\hspace{1em}
}
}
\end{figure}

Notice that the splitting of $A$ into $A+L$ and $A-L$, and therefore also the spectra of $\AA|_{\HH_+}$ and $\AA|_{\HH_-}$,
depend only on $L$; they do not depend on $\theta$, which measures the relative strengths of the bias~$B$ and coupling~$C$.  What changes with $\theta$ are the relative amplitudes of the components of the hybrid fields,
as seen from the definitions of $\HH_+$ and $\HH_-$.  The energy (square norm) of a hybrid state is divided between the two graphs:
\begin{eqnarray}
  && \hspace{1.3em} \| (u_1,u_2) \|^2 \,=\, \|u_1\|^2 + \|u_2\|^2 \\
  && \renewcommand{\arraystretch}{1.2}
\left\{
  \begin{array}{l}
    \|u_1\|^2 = \cos^2(\theta/2)\|u\|^2 \\
    \|u_2\|^2 = \sin^2(\theta/2)\|u\|^2    
  \end{array}
\right.
\quad \text{for }\; (u_1,u_2)=\left( \cos(\textstyle\theta/2)\,u\,,\; e^{-i\phi}\!\sin(\theta/2)\,u \right)\in\HH_+ \\
  && \renewcommand{\arraystretch}{1.2}
\left\{
  \begin{array}{l}
    \|u_1\|^2 = \sin^2(\theta/2)\|u\|^2 \\
    \|u_2\|^2 = \cos^2(\theta/2)\|u\|^2    
  \end{array}
\right.
\quad \text{for }\; (u_1,u_2)=\left( -e^{i\phi}\!\sin(\textstyle\theta/2)\,u\,,\; \cos(\theta/2)\,u \right)\in\HH_-\,.
\end{eqnarray}
Figure~\ref{fig:theta} illustrates the relation between the relative strengths of the bias and coupling and the relative amplitudes of the components of the hybrid states.  When the two graphs are coupled but no bias is imposed ($\theta=\pi/2$), the energy of a hybrid state is equally partitioned between the two graphs.  If, in addition, $\phi=0$, $\HH_+$ and $\HH_-$ consist of the symmetric and antisymmetric states:
\begin{equation*}
\renewcommand{\arraystretch}{1.3}
\left.
  \begin{array}{l}
    \HH_+ \,=\, \left\{ (u,u) \,:\, u\in\HH \right\} \\
    \HH_- \,=\, \left\{ (u,-u) \,:\, u\in\HH \right\}
  \end{array}
\right.
\quad
\text{[\,$\theta = \pi/2$ (no bias) and $\phi=0$\,].}
\end{equation*}
 On the other extreme, $\theta=0$ corresponds to no coupling, so $\HH_+=\HH\oplus\{0\}$ and $\HH_-=\{0\}\oplus\HH$.

The operator $L$ is reminiscent of the {Rabi frequency}; if $L>0$, it is determined through the bias and the coupling by $L^2 = B^2 + CC^*$.  If $L=\lambda I$, the number $2\lambda$ is the width of the spectral splitting of the two decoupled spaces $\HH_+$ and $\HH_-$ of hybrid states.

\subsection{Spectrally embedded eigenfunctions}\label{sec:eigenfunctions}

The coupling-decoupling technique of sec.~\ref{sec:decoupling} can be used to create locally perturbed periodic graph operators with an embedded eigenvalue whose eigenfunction has unbounded support.
By specializing the operator $L$ to a multiple $\lambda_0 I$ of the identity, the spectra of the hybrid systems can be shifted at will.  To create an embedded eigenvalue for the operator $\mathcal A$ for some $\lambda_0$, it suffices that $A$ possess an eigenvalue, embedded or~not.

\begin{theorem}\label{thm:embedded}
  Let $A$ be a periodic self-adjoint operator on a combinatorial or metric graph $\Gamma$ and $V$ a compactly supported self-adjoint perturbation of $A$, and suppose that $u\in L^2(\Gamma)$ satisfies $(A+V)u = \lambda u$.  Let $\lambda_0\in\RR$ be such that $\lambda+2\lambda_0\in\sigma_c(A)$, where $\sigma_c(A)$ denotes the continuous spectrum of $A$.  For each $\theta\in\RR$ and $\phi\in\RR$, $\lambda+\lambda_0$ is an embedded eigenvalue of each the self-adjoint operators
\begin{equation}\label{perturbed}
  \mathcal A+ \mathcal V_1
  \;=\; \renewcommand{\arraystretch}{1.4}
\left[
  \begin{array}{cc}
    A + \lambda_0\cos\theta\;I & e^{i\phi}\lambda_0\sin\theta\; I \\
    e^{-i\phi}\lambda_0\sin\theta\;I & A-\lambda_0\cos\theta\; I
  \end{array}
\right]
+ \renewcommand{\arraystretch}{1.4}
\left[
  \begin{array}{cc}
    V & 0 \\
    0 & V
  \end{array}
\right]
\end{equation}
and
\begin{equation}\label{perturbed}
  \mathcal A+ \mathcal V_2
  \;=\; \renewcommand{\arraystretch}{1.4}
\left[
  \begin{array}{cc}
    A + \lambda_0\cos\theta\;I & e^{i\phi}\lambda_0\sin\theta\; I \\
    e^{-i\phi}\lambda_0\sin\theta\;I & A-\lambda_0\cos\theta\; I
  \end{array}
\right]
+ \renewcommand{\arraystretch}{1.4}
\left[
  \begin{array}{cc}
    \cos^2\frac{\theta}{2}\;V & \frac{1}{2}e^{i\phi}\sin\theta\; V \\
    \frac{1}{2}e^{-i\phi}\sin\theta\; V & \sin^2\frac{\theta}{2}\; V
  \end{array}
\right]
\end{equation}
in $L^2(\Gamma)\oplus L^2(\Gamma)$.  In each case, the eigenfunction corresponding to $\lambda+\lambda_0$ is
\begin{equation}\label{eigenfunction}
  \tilde u \,=\, \left( \cos\textstyle\frac{\theta}{2}\;u\,, e^{-i\phi}\sin\frac{\theta}{2}\;u \right)\,,
\end{equation}
that is, $(\mathcal A + \mathcal V_{1,2})\tilde u = \lambda\tilde u$.

In particular, if $\Gamma$ is a combinatorial graph and $u$ has unbounded support, then $\tilde u$ is an eigenfunction (bound state) of unbounded support for an embedded eigenvalue of the self-adjoint operators $\mathcal A + \mathcal V_{1,2}$ on a combinatorial graph whose vertex set is $\mathrm{vert}(\Gamma)\mathring\cup \mathrm{vert}(\Gamma)$, where $\mathrm{vert}(\Gamma)$ is the vertex set of $\Gamma$.  Note that, in this case, $L^2(\Gamma)=\ell^2(\mathrm{vert}(\Gamma))$.
\end{theorem}

\noindent
{\bfseries Remark.} The perturbation $\mathcal V_2$ vanishes (acts as the zero operator) on the subspace $\mathcal H_-$, and therefore does not affect the extended states associated with $\mathcal H_-$.
The bound state $\tilde u$ is in the subspace $\mathcal H_+$, on which $\mathcal V_2$ acts as a local perturbation of $\mathcal A$.  
The perturbation $\mathcal V_1$, on the other hand, affects the action of $\mathcal A$ in both subspaces $\mathcal H_\pm$.  Since $V$ is a local perturbation, it does not modify the continuous spectrum of $A$, and thus neither does $\mathcal V_1$ modify the continuous spectrum of $\mathcal A$.  Thus the eigenvalue $\lambda+\lambda_0$, which is by construction within the continuum of $\mathcal A$, is also within the continuum of $\mathcal A+ \mathcal V_1$.

\begin{proof}
  Assume that, for some $u\in L^2(\Gamma)$, $(A+V)u = \lambda u$ and $\lambda+2\lambda_0\in\sigma_c(A)$, and put $L=\lambda_0 I$.  Then
$(A+L+V-(\lambda+\lambda_0))u=0$ and $\lambda+\lambda_0\in\sigma_c(A)-\lambda_0=\sigma_c(A-L)$.
Since $V$ is a local graph operator, $\sigma_c(A-L+V)=\sigma_c(A-L)$, and thus $\lambda+\lambda_0\in\sigma_c(A-L+V)$.
This means that $\lambda+\lambda_0$ is an embedded eigenvalue of the operators
\begin{equation*}
  \tilde{\mathcal A}_1 \,=\, \tilde{\mathcal A}+\mat{1.3}{V}{0}{0}{V}
  \,=\, \mat{1.3}{A+L+V}{0}{0}{A-L+V}
\end{equation*}
and
\begin{equation*}
  \tilde{\mathcal A}_2 \,=\, \tilde{\mathcal A}+\mat{1.3}{V}{0}{0}{0}
  \,=\, \mat{1.3}{A+L+V}{0}{0}{A-L}
\end{equation*}
in $L^2(\Gamma)\oplus L^2(\Gamma)$ with eigenfunction $[u,0]^t$.
Therefore $\lambda+\lambda_0$ is an embedded eigenvalue of $\mathcal{U}\tilde{\mathcal A}_{1,2}\mathcal{U}^{-1}$, where $\mathcal U$ is the unitary operator defined by (\ref{U}).  One computes using (\ref{AUUA1}) that $\mathcal{U}\tilde{\mathcal A}_{1,2}\mathcal{U}^{-1}$ is equal to ${\mathcal A}+{\mathcal V_{1,2}}$ in the theorem, and a corresponding eigenfunction is $\mathcal{U}[u,0]^t = \tilde u$.
\end{proof}

This theorem allows one to use any (typically non-embedded) eigenvalue of a locally perturbed periodic operator on a combinatorial graph $\Gamma$ to construct an embedded eigenvalue for an operator on another graph, namely the union of two disjoint copies of $\Gamma$.  Non-embedded eigenvalues whose eigenfunctions have unbounded support and exponential decay are commonplace for locally defective periodic structures; a construction for graphs is given below.  This observation, together with Theorem~\ref{thm:embedded} yields

\begin{corollary}
  There exist self-adjoint $n$-periodic ($n\in\ZZ$) finite-degree combinatorial graph operators that admit localized self-adjoint perturbations possessing an embedded eigenvalue whose eigenfunction has unbounded support and exponential decay.
\end{corollary}

The following discussion shows how to construct, for a simple class of graph operators, a non-embedded eigenvalue whose eigenfunction has unbounded support and exponential decay.  Let $A$ be a degree-$d$ ($d<\infty$) $n$-periodic difference operator on a graph $\Gamma$ whose fundamental domain $W$ consists of a single vertex.  The graph $\Gamma$ can be identified with the integer lattice $\ZZ^n$.  The perturbation $V$ will be a multiplication operator with support at a single vertex.

First consider the forced equation
\begin{equation}\label{forced}
  (A-\lambda I)u = \delta\,,
\end{equation}
in which $\delta(0)=1$ and $\delta(g)=0$ for all nonzero $g\in\ZZ^n$.  Application of the $z$-transform gives the scalar equation
\begin{equation*}
  (\hat A(z) - \lambda)\hat u(z) = 1\,,
\end{equation*}
in which $\hat A(z)=\sum_{|g|\leq d} A_{g}z^{g}$ is a Laurent polynomial in $z\in\CC^n$.
Assuming that $\lambda\in\RR\setminus\sigma(A)$, the number $\hat A(z) - \lambda$ is nonzero for all $z\in\TT^n$ and the function
\begin{equation*}
  \hat u(z) = \frac{1}{\hat A(z) - \lambda}
\end{equation*}
is bounded on $\TT^n$.  By the Fourier inversion theorem, $\hat u$ is the $z$-transform of a function $u$ in $\ell^2(\Gamma)$, which satisfies $(A-\lambda I)u = \delta$.  The solution $u$ has bounded support if and only if $\hat u(z)$ is a Laurent {polynomial} in $z$.
Assuming that $A$ is not a multiplication operator (there are interactions between vertices), $\hat A(z)$ is non-constant.  Moreover, since $A$ is self-adjoint, the coefficients of $\hat A(z)$ satisfy $A_{-g}=\overline{A_g}$, and thus $\hat A(z) - \lambda$ has at least two nonzero terms.  It follows that $\hat A(z) - \lambda$ vanishes at some $z\in(\CC^*)^n$ so that $\hat u(z)$ cannot be a Laurent polynomial.  Thus $u$ has unbounded support.
Since $\hat u(z)$ is analytic in a complex neighborhood of $\TT^n$, $u$ is an exponentially decaying function of $g\in\ZZ^n$.

The Floquet inversion theorem gives $u(0)$ as the average of $\hat u(z)$ over the $n$-torus:
\begin{equation*}
  u(0) \,=\, \frac{1}{(2\pi)^n}\int_\TT \hat u(e^{ik_1},\dots,e^{ik_n})\, dk_1\cdots dk_n
    \,=\, \frac{1}{(2\pi)^n}\int_\TT \big( \hat A(e^{ik_1},\dots,e^{ik_n}) -\lambda \big)^{-1} dk_1\cdots dk_n \,\not=\, 0\,.
\end{equation*}
The value $u(0)$ is real and nonzero.
The reason is the identity $\overline{\hat A(\bar z^{-1})}=\hat A(z)$ coming from the self-adjointness of $A$, which makes $\hat A(z)$ real valued on $\TT^n$.  Since $\hat A(z)\!-\!\lambda$ is real, non-vanishing, and continuous at each $z\in\TT^n$, the integrand is of one sign.

Define the multiplication operator $V$ on $\ell^2(\Gamma)$ by
\begin{equation*}
  (Vf)(g) = \renewcommand{\arraystretch}{1.1}
\left\{
  \begin{array}{ll}
    -u(0)^{-1} & \text{if $g=0$} \\
    0 & \text{if $g\not=0$}
  \end{array}
\right.
\qquad\text{for } g\in\ZZ^n\,.
\end{equation*}
By this definition, $-Vu=\delta$, and by (\ref{forced}), one obtains
\begin{equation*}
  (A+V)u = \lambda u\,,
\end{equation*}
so that $u$ is an eigenfunction of $A+V$ with unbounded support and exponential decay.

\subsection{Generalization to several coupled graphs}\label{sec:several} 

The construction of two spaces of decoupled hybrid states can be generalized to $m$ spaces, leading to embedded eigenvalues in systems of several coupled graph operators.

One does this by generalizing the matrix
\begin{equation*}
  K = \mat{1.1}{\cos\theta}{e^{i\phi}\sin\theta}{e^{-i\phi}\sin\theta}{-\cos\theta}
\end{equation*}
of biases and couplings to any $m\times m$ Hermitian matrix $K$ whose entries will serve as coupling coefficients among $m$ identical graphs.  The hybrid states are defined through the columns of a unitary matrix $U$ that diagonalizes~$K$, thus generalizing the matrix
\begin{equation*}
U =      \renewcommand{\arraystretch}{1.5}
\left[\!
  \begin{array}{cc}
    \cos(\theta/2) & -e^{i\phi} \sin(\theta/2) \\
    e^{-i\phi}\sin(\theta/2) & \cos(\theta/2)
  \end{array}
\!\right]
\end{equation*}
from the two-system case.  The operators on the $m$ hybrid state spaces are of the form $A+\lambda_i L$, where the $m$ real numbers $\lambda_i$ are the eigenvalues of $K$.  All this is made precise below.

\smallskip
Let a self-adjoint operator $A$ in a Hilbert space $\mathcal H$ be given, and consider $m$ identical copies of the system $(\HH, A)$ coupled through multiples of a single bounded self-adjoint operator $L:\HH\to\HH$.  This results in an operator $\mathcal A$ in the direct sum $\oplus_{j=1}^m\HH$:
\begin{eqnarray*}
  &&\AA\;:\; \bigxor_{j=1}^m \mathrm{dom}(A) \to \bigxor_{j=1}^m \HH \\
  && \AA(v_1,\dots v_m) = (w_1,\dots w_m),
  \quad
  w_i \;=\; Av_i \,+\, \sum_{j=1}^m \Centry_{ij} Lv_j\,.
\end{eqnarray*}
The $m\times m$ matrix of coupling coefficients $\C=(\Centry_{ij})$ is Hermitian, which makes $\mathcal A$ self-adjoint.  The ``self-couplings" provided by the diagonal entries of $\C$ can be thought of as modifications of the operator $A$ on each copy of $(\HH,A)$, generalizing the bias $B$ from before.
By identifying $\oplus_{j=1}^m\HH$ with a tensor product,
\begin{equation*}
  \bigxor_{j=1}^m \HH \;\cong\; \CC^m \tensor \HH\,,
\end{equation*}
the block-matrix form of $\mathcal A$ is written conveniently as
\begin{equation*}
  \AA \;=\; I_m \tensor A \,+\, \C \tensor L\,,
\end{equation*}
in which $I_m$ is the $m\times m$ identity matrix.

The operator $\mathcal A$ can be block-diagonalized.  Let $U=(\gamma_{ij})$ be the unitary $m\times m$ matrix that conjugates $\C$ into a diagonal matrix $\Lambda$ of real eigenvalues $\lambda_j$ of $\C$,
\begin{equation*}
  \C\, U \,=\, U\Lambda\,.
\end{equation*}
The tensor product ${\mathcal U}=U\tensor I$, where $I$ is the identity operator on $\mathcal H$, is an $m\times m$ block matrix whose blocks are the multiples $\gamma_{ij}I$ of the identity.  It is a unitary operator on $\CC^m \tensor \HH$ that decomposes $\mathcal A$ into $m$ subsystems:
\begin{equation*}
  \left( I_m\tensor A \,+\, \C\tensor L \right) U\tensor I \;=\; U\tensor I \left( I_m\tensor A + \Lambda\tensor L \right),
\end{equation*}
or, more concisely,
\begin{equation}\label{AUUA2}
  \AA\,{\mathcal U} \,=\, {\mathcal U}\tilde\AA\,.
\end{equation}
Here, $\tilde\AA=\left( I_m\tensor A + \Lambda\tensor L \right)$ is a block-diagonal matrix whose diagonal blocks are modifications of $A$:
\begin{equation}\label{tildeA}
  \tilde\AA \,=\, 
  \renewcommand{\arraystretch}{1.1}
\left[
  \begin{array}{cccc}
    A+\lambda_1L & 0 & \cdots & 0 \\
    0 & A+\lambda_2L & \ddots & \vdots \\
    \vdots & \ddots & \ddots & \vdots \\
    0 & \cdots & \cdots & A+\lambda_mL \\
  \end{array}
\right]\,.
\end{equation}

\smallskip
Now let $A$ and $L$ be periodic difference operators of finite degree on a combinatorial graph $\Gamma$ with $\ZZ^n$ translational symmetry, and let $\hat A(z)$, $\hat\AA(z)$, {\itshape etc.}, be the spectral representations of the corresponding operators under the Floquet transform.  
From (\ref{AUUA2}), one obtains
\begin{equation*}
  \big( \hat\AA(z) - \lambda{\mathcal I}\, \big)\, {\mathcal U} \,=\, {\mathcal U} \big( \hat{\tilde{\mathcal A}}(z) - \lambda\,{\mathcal I}\, \big)\,,
\end{equation*}
in which $\mathcal I$ is the identity operator on $\oplus_{j=1}^m\HH$.
The operator $\hat{\tilde{\mathcal A}}(z)$ has a block-diagonal form obtained by replacing $A$ and $L$ with their Floquet transforms in (\ref{tildeA}).  Thus the Floquet surface of $\mathcal A$ is reducible for each energy $\lambda$:
\begin{equation*}
  0 \,=\, \det\big( \hat\AA(z) - \lambda\,{\cal I}\, \big)
  \,=\, \det\big( \hat{\tilde{\mathcal A}}(z) - \lambda\,{\cal I}\, \big) 
  \,=\, \prod_{i=1}^m \det \big( \hat A(z) + \lambda_i\hat L(z) - \lambda\, I \big)\,.
\end{equation*}
One can then construct embedded eigenvalues for the operator ${\mathcal A}$ in the combinatorial graph whose vertex set is the union of $m$ disjoint copies of $\Gamma$ by generalizing the procedure in sec.~\ref{sec:eigenfunctions}.

\section{Embedded eigenvalues in quantum graphs}\label{sec:quantum}

If $(\Gamma,A)$ is a quantum graph and $\HH=L^2(\Gamma)$, the system $(\HH\oplus\HH,\mathcal A)$ constructed in section~\ref{sec:decoupling} does not define a quantum graph because of the direct coupling between vertices of the two copies of $\Gamma$.  The construction can be modified by realizing the coupling through additional edges connecting the two copies of $\Gamma$.  The main result of this section is that  {\em there exist self-adjoint $n$-periodic finite-degree quantum graphs that admit localized self-adjoint perturbations that possess an embedded eigenvalue whose eigenfunction has unbounded support.}

First, a general procedure for constructing embedded eigenvalues is developed in section~\ref{sec:quantumcoupling}.
It involves ``decorating" a given graph by periodically attaching dangling edges, which creates gaps in the spectrum that depend on the condition at the free vertex; see~\cite{SchenkerAizenman2000} for a proof of this phenomenon for combinatorial graphs.  When two identical copies of the decorated graph are connected at the free vertices, the resulting graph decouples into even and odd states whose spectra are equal to those for the free-endpoint (Neumann) and clamped-endpoint (Dirichlet) conditions imposed on the free vertices of the decorated graph.  One then tries to construct an eigenvalue in a spectral gap of the even (odd) states that lies in a band of the odd (even) states to produce an eigenvalue that is embedded in the spectrum of the full system.
A full proof for a specific 2D graph is presented in section~\ref{sec:quantumeigenvalues} (Theorem~\ref{thm:quantum}).

\subsection{Coupling two quantum graphs by edges and (anti)symmetric states}\label{sec:quantumcoupling}

\begin{figure}  
\floatbox[{\capbeside\thisfloatsetup{capbesideposition={right,top},capbesidewidth=7cm}}]{figure}[\FBwidth]
{\caption{\small
{\bfseries Left.} A fundamental domain $\tilde W$ of a doubly periodic metric graph $\tilde\Gamma$ consists of the solid vertices and the edges shown.  It is built from two copies of a fundamental domain $W$ of a graph $\Gamma$ (graphene in this example) connected by an edge $e_0$ parameterized by $[-1/2,1/2]$.
{\bfseries Right.}  The graph $\Gamma_*$ is half of $\tilde\Gamma$.  Its fundamental domain $W_*$ is half of $\tilde W$, and has one dangling edge $e'_0$ parameterized by $[0,1/2]$.}
\label{fig:QuantumGraph2}}
{
\raisebox{4ex}{\scalebox{0.235}{\includegraphics{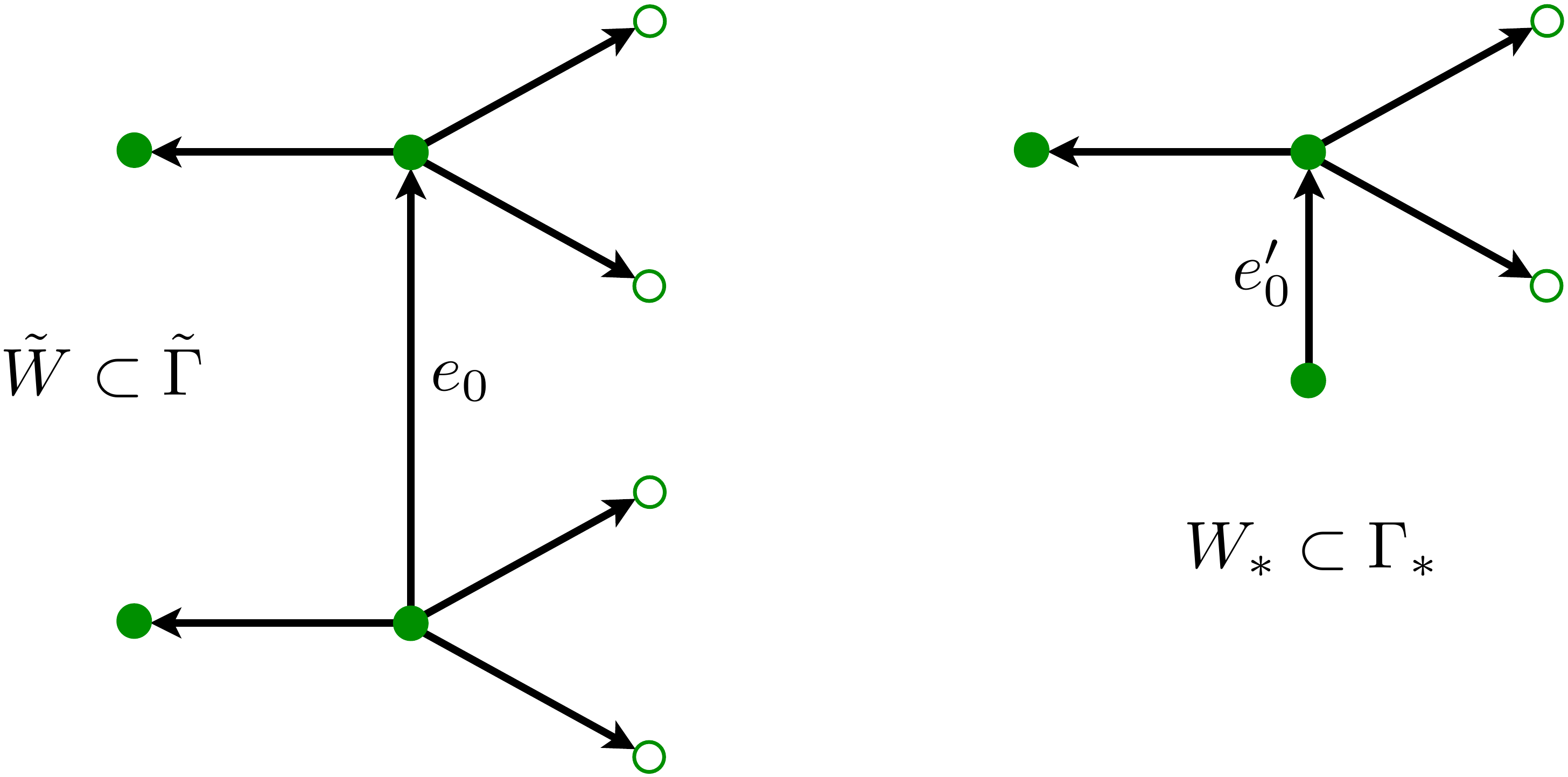}}
\hspace{2em}
}
}
\end{figure}

Let $(\Gamma,A)$ be an $n$-periodic quantum graph, and let $\tilde\Gamma$ be the graph obtained by connecting two identical copies $(\Gamma_1,A_1)$ and $(\Gamma_2,A_2)$ of $(\Gamma,A)$ by edges that connect vertices in $\Gamma_1$ to the corresponding ones in $\Gamma_2$, to obtain a periodic metric graph $\tilde\Gamma$.  Endow $\tilde\Gamma$ with a periodic operator $\tilde A$ given by $A$ on the edges of $\Gamma_1$ and $\Gamma_2$ and by $-d^2/dx^2+q(x)$ on the connecting edges.  A fundamental domain $\tilde W$ of the quantum graph $(\tilde\Gamma,\tilde A)$ consists of two copies of a fundamental domain $W$ of $\Gamma$ connected by, say, just one edge $e_0$ for simplicity.  Fig.~\ref{fig:QuantumGraph2} depicts the case that $\Gamma$ is the hexagonal graph of graphene.  Let $e_0$ be identified with the $x$-interval $[-\onehalf,\onehalf]$ and $q$ be symmetric.

A local perturbation of $\tilde A$ analogous to that in Example~3 consists of a constant potential $V_0$ applied only to the edge $e_0$ in the fundamental domain $\tilde W$, but not to any of the translates of $e_0$; call this potential $\tilde V$:
\begin{equation*}
  (\tilde Vu)(p) =
  \renewcommand{\arraystretch}{1.1}
\left\{
  \begin{array}{ll}
    V_0 & \text{if }\, p\in e_0\subset\tilde W, \\
    0 & \text{otherwise}.
  \end{array}
\right.
\end{equation*}

The operator $\tilde A$ is reduced by the decomposition $L^2(\tilde\Gamma)= \HH_+\oplus\HH_-$, where $\HH_+$ ($\HH_-$) is the space of functions symmetric (anti-symmetric) with respect to reflection about the center of $e_0$ and its translates and switching of $\Gamma_1$ and $\Gamma_2$.  Thus the spectrum of $\tilde A$ is the union of the spectra of $\tilde A$ restricted to the spaces $\HH_\pm$.  The restriction $\tilde A|_{\HH_+}$ is identified with the quantum graph $(\Gamma_*,A_+)$, where $\Gamma_*$ is ``half" of $\tilde\Gamma$---its fundamental domain $W_*$ consists of $W$ plus half of $e_0$ dangling from one vertex of $W$ (Fig.~\ref{fig:QuantumGraph2}).  Call this edge $e_0'$; it is coordinatized by the interval $[0,\onehalf]$.  The action of the operator $A_+$ coincides with that of $\tilde A$ on each edge, but its domain is subject to the Neumann boundary condition $u'(0)=0$ on the free vertex of~$e_0'$.  Similarly, the restriction $\tilde A|_{\HH_-}$ is identified with the quantum graph $(\Gamma_*,A_-)$, where $A_-$ is subject to the Dirichlet boundary condition $u(0)=0$ on~$e_0'$.
Because $V_0$ is symmetric about the center point of $e_0$, the decomposition of $\tilde A+\tilde V$ by the spaces $\HH_\pm$ persists.

The objective, for any given quantum graph $(\Gamma,A)$, is to find an interval $I$ contained simultaneously in a spectral band of $A_+$ and in a spectral gap of $A_-$ (or vice-versa) and such that a localized perturbation creates a (non-embedded) eigenvalue for $A_-$ in $I$ and thus an embedded eigenvalue for $\tilde A$.  One expects this procedure to be generically possible because resonant excitement of the dangling edge $e'_0$ creates gaps in the spectrum around the eigenvalues of $e'_0$ with Dirichlet condition at the connecting vertex and Dirichlet or Neumann condition at the free vertex \cite{SchenkerAizenman2000}, \cite[Ch.~5]{BerkolaikoKuchment2013} (although the eigenvalue itself has infinite multiplicity and thus remains in the spectrum).  This was demonstrated in Example~3 in the case of 1D periodicity and will be seen again in sec.~\ref{sec:quantumeigenvalues}.

\smallskip
The rest of this subsection shows how to create a non-embedded eigenvalue for $A_+$ or $A_-$, assuming that $q(x)=0$ on the edges connecting $\Gamma_1$ to $\Gamma_2$.  The procedure can be generalized to nonzero symmetric~$q$.  Consider first the forced problem
\begin{equation*}
  (A_\pm-\lambda I) u = f\,,
\end{equation*}
in which $f$ vanishes everywhere on $\Gamma_*$ except on the dangling edge $e_0'$ in the fundamental domain $W_*$, where for $x\in e_0'$, $f(x)=\cos\nu x$ for the graph $(\Gamma_*,\,\HH_+)$ (Neumann) and $f(x)=\frac{1}{\nu}\sin\nu x$ for the graph $(\Gamma_*,\,\HH_-)$ (Dirichlet) for some $\nu>0$. (Note that $e_0'$ is being identified with $[0,1/2]$.)

In the Neumann case, assume that $\mu^2=\lambda\not\in\sigma(A_+)$, so that $(A_+-\lambda I)u=f$ has a unique solution $u\in\mathrm{dom}(A_+)$.  The solution $u$ satisfies $-u''-\mu^2 u=0$ on each edge except the dangling edge $e_0'$ in $W_*$, where it satisfies
\begin{eqnarray*}
  && -u''-\mu^2 u \,=\, \cos\nu x \\
 && \hspace{0.8em} u'(0) = 0\,.
\end{eqnarray*}
The solution is
\begin{equation}\label{uoneN}
  u(x) \,=\, \frac{1}{\nu^2-\mu^2}\cos\nu x \,+\, K_N(\mu,\nu)\cos\mu x\,
  \qquad
  \text{for $x\in e_0'\subset W_*$}
\end{equation}
for some constant $K_N(\mu,\nu)$.
One has \,$-V_0u(x) = \cos\nu x=f(x)$\, for some $V_0\in\RR$ if and only if $K_N(\mu,\nu)=0$.  In this case,
\begin{equation*}
  V_0 = \mu^2-\nu^2\,
\end{equation*}
and hence the equation
\begin{equation*}
  (A_++V)u = \mu^2 u,
  \qquad
  \text{(\,if $K_N(\mu,\nu)=0$\,)}
\end{equation*}
holds, where $V$ is the multiplication operator
\begin{equation*}
  (Vu)(p) =
  \renewcommand{\arraystretch}{1.1}
\left\{
  \begin{array}{ll}
    V_0 & \text{if $p\in e_0'\subset W_*$}\\
    0 & \text{otherwise}\,,
  \end{array}
\right.
\end{equation*}
so that $u$ is an eigenfunction of $A_++V$ with eigenvalue $\mu^2$.
Given $\mu$, one would like to determine $\nu$ such that $K_N(\mu,\nu)=0$, and thus the perturbation $V$ that creates an eigenvalue of $A_++V$.

Under the Floquet transform,
\begin{equation*}
  (\hat A_+(z) - \lambda I)\hat u(p,z) \,=\, \hat f(p,z),
  \qquad (p\in\Gamma_*\,,\; z\in\ZZ^n)
\end{equation*}
in which $\hat f(p,z)$ is $z$-quasi-periodic on $\Gamma_*$ and independent of $z$ for all $p$ in the fundamental domain $W_*$ because $f$ is supported in $W_*$:
\begin{equation*}
  \hat f(gp,z) \,=\, f(p)z^g
  \qquad
  \text{for all $p\in W_*$.}
\end{equation*}
One obtains
\begin{equation}\label{uhat}
  \hat u(p,z) \,=\, (\hat A_+(z) - \lambda I)^{-1} f(p)
  \qquad
  \text{for $p\in W_*$.}
\end{equation}
By the inverse Floquet transform,
\begin{equation}\label{up}
  u(p) \,=\, \frac{1}{(2\pi)^n} \int_{\TT^n} \hat u(p;e^{ik_1},\dots,e^{ik_n}) dk_1\dots dk_n
  \qquad \text{for $p\in W_*$}.
\end{equation}
For any $z\in(\CC^*)^n$, the operator $\hat A_+(z)$ is the restriction of $A_+$ to $z$-quasi-periodic functions on $W_*$ with the Neumann boundary condition on the free vertex of the dangling edges, and thus, on $e_0'$, $\hat u$ satisfies
\begin{eqnarray*}
  &&-\hat u'' - \mu^2\hat u \,=\, \cos\mu x
  \qquad
  (x\in e_0'), \\
  && u'(0) = 0,
\end{eqnarray*}
in which $k=(k_1,\dots,k_n)$ and $z=(e^{ik_1},\dots,e^{ik_n})$, so the solution is
\begin{equation}\label{uhate}
  \hat u(x,z) \,=\, \frac{\cos\nu x}{\nu^2-\mu^2} \,+\, \hat K_N(\mu,\nu;k)\cos\mu x
  \qquad
  (x\in e_0').
\end{equation}
Because of (\ref{uhate}), (\ref{uoneN}) and (\ref{up}), the coefficient $K_N(\mu,\nu)$ (\ref{uoneN}) of $u$ on the dangling edge $e_0'$ in $W_*$ is
\begin{equation}\label{KN}
  K_N(\mu,\nu) \,=\, \frac{1}{(2\pi)^n} \int_{\TT^n} \hat K_N(\mu,\nu;k_1,\dots,k_n)dk_1\dots dk_n\,.
\end{equation}

Still assuming $\lambda=\mu^2\not\in\sigma(A_+)$, the solution $u$ {\em decays exponentially} by standard theorems of Fourier transforms:
In each $g$-translate of $e'_0$ ($g\in\ZZ^n$), $u$ has the form $K_N(\mu,\nu;g)\cos\mu x$.
The $z$-transform of $K_N(\mu,\nu;g)$, namely $\hat K_N(\mu,\nu;\cdot)$ as a function of $z=(e^{ik_1},\dots,e^{ik_2})$ in $(\CC^*)^n$, is analytic in a neighborhood of the torus $\TT^n$ because, by (\ref{uhat}), $\hat u(p,\cdot)$ is.  Therefore the coefficient $K_N(\mu,\nu;g)$ is exponentially decaying as a function of $g$.  Similarly, on each non-dangling edge $ge$, with $e$ and edge in $W_*$, $u(x)=C(g)\cos\mu x + D(g)\cos\mu x$, and one finds that these coefficients are also exponentially decaying in $g$.  Thus $u$ itself decays exponentially.

To show that $u$ has {\em unbounded support}, one has to prove that the Floquet transform $\hat u(p,z)$ ($p\in\Gamma_*$ and $z\in\ZZ^n$) is not a Laurent polynomial in $z$.
This is achieved by arguments similar to those in section~\ref{sec:eigenfunctions}.  In the quantum-graph case, one first reduces the differential operator $\hat A_+(z)-\lambda I$ to a matrix $\hat {\mathfrak A}_+(z,\lambda)$ acting on the vector of coefficients $\hat K$, $\hat C$, {\itshape etc.}, representing the solution $\hat u$ on the edges of $W_*$, and then shows that $\det(\hat {\mathfrak A}_+(z,\lambda))$ must vanish on some nonempty surface in $(\CC^*)^n$.  It follows generically that the coefficients $\hat K$, $\hat C$, {\itshape etc.}, have poles in $z\in(\CC^*)^n$ and are therefore not Laurent polynomials.  One has only to check that the vector representing the forcing is not in the range of the matrix $\hat {\mathfrak A}_+(z,\lambda)$.  This process is carried out for a particular quantum graph in the next subsection.

Analogous arguments hold for the Dirichlet problem for the operator $A_-$.  
The locally forced problem is
\begin{eqnarray*}
  && -u''-\mu^2 u \,=\, {\textstyle\frac{1}{\nu}}\sin\nu x \\
 && \hspace{0.8em} u(0) = 0\,,
\end{eqnarray*}
and its solution is
\begin{equation}\label{uoneD}
  u(x) \,=\, \frac{{\textstyle\frac{1}{\nu}}\sin\nu x}{\nu^2-\mu^2} \,+\, K_D(\mu,\nu){\textstyle\frac{1}{\mu}}\sin\mu x\,
  \qquad
  \text{for $x\in e_0'\subset W_*$}\,.
\end{equation}
The Floquet transform $\hat u$ satisfies
\begin{eqnarray*}
  &&-\hat u'' - \mu^2\hat u \,=\, \frac{1}{\mu}\sin\mu x
  \qquad
  (x\in e_0'), \\
  && u(0) = 0, \\
  && \hat u(x,z) \,=\, \frac{\frac{1}{\nu}\sin\nu x}{\nu^2-\mu^2} \,+\, \hat K_D(\mu,\nu;k){\textstyle\frac{1}{\mu}}\sin\mu x
  \qquad
  (x\in e_0').
\end{eqnarray*}
An expression analogous to (\ref{KN}) holds for $K_D(\mu,\nu)$.

\subsection{Embedded eigenvalues for a 2D quantum graph}\label{sec:quantumeigenvalues}

The procedure for creating embedded eigenvalues with unbounded support outlined in the previous section is carried out for a specific quantum graph, namely, a two-dimensional version of Example~3.
 
\begin{figure}  
\floatbox[{\capbeside\thisfloatsetup{capbesideposition={right,top},capbesidewidth=6.6cm}}]{figure}[\FBwidth]
{\caption{\small This quantum graph $(\tilde\Gamma,\tilde A)$ consists of two copies of the square planar grid connected by vertical edges.  The operator $\tilde A$ of the graph is $-d^2/dx^2$ on each edge except on one vertical edge (bold), where it is $-d^2/dx^2+V_0$.
This local defect results in embedded eigenvalues with eigenfunctions that have unbounded support and decay exponentially.
}
\label{fig:QuantumGraph3}}
{
\raisebox{4ex}{\scalebox{0.3}{\includegraphics{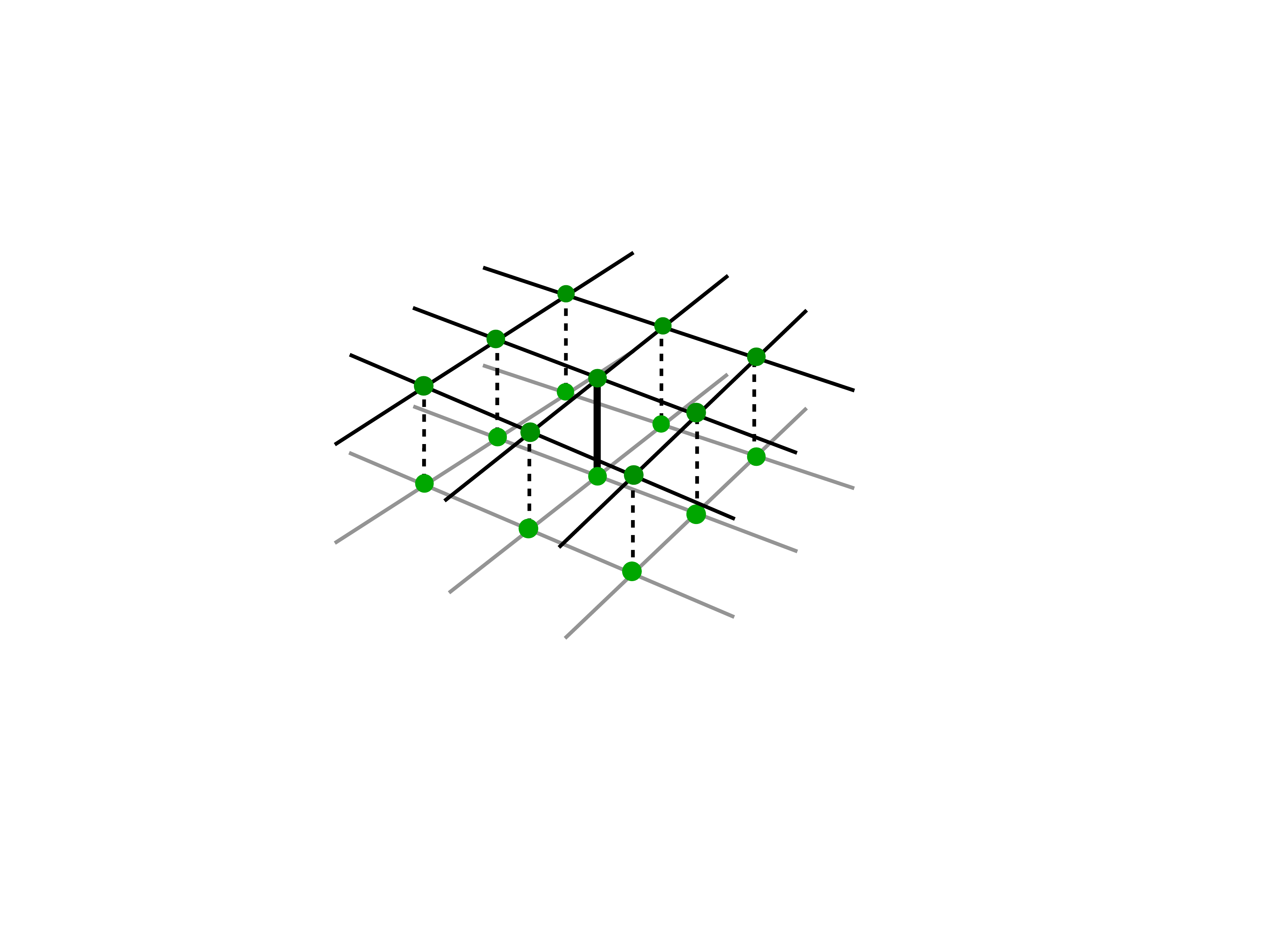}}
\hspace{1em}
}
}
\end{figure}

   Let $\tilde\Gamma$ be the metric graph whose vertex set is two stacked copies of the integer lattice $\ZZ^2$, or, more concretely, the integer triples $(\ell_1,\ell_2,\ell_3)$ with $\ell_3$ equal to $0$ or $1$, and whose edges connect adjacent vertices along the coordinate directions (Fig.~\ref{fig:QuantumGraph3}).  Let $\tilde A$ be the self-adjoint operator acting by $-d^2/dx^2$ on each edge and whose domain is subject to continuity and the zero-flux (a.k.a.~Neumann) condition at the vertices.
Let a localized potential $\tilde V$ be defined by a multiplication operator that vanishes on each edge except one of the vertical edges connecting the two copies of $\ZZ^2$, on which $\tilde V$ is equal to a constant $V_0$.  
 
\begin{theorem}\label{thm:quantum}
For suitable values of $V_0$, the locally perturbed periodic quantum graph $(\tilde\Gamma,\tilde A+\tilde V)$ (Fig.~\ref{fig:QuantumGraph3}) admits an embedded eigenvalue whose eigenfunction is exponentially decaying, has unbounded support, and is either symmetric or anti-symmetric with respect to reflection about the plane midway between the two copies of $\ZZ^2$ in~$\tilde \Gamma$. 
\end{theorem} 

The remainder of this section is a proof of this theorem.

\begin{figure}  
\floatbox[{\capbeside\thisfloatsetup{capbesideposition={right,top},capbesidewidth=6.8cm}}]{figure}[\FBwidth]
{\caption{\small A fundamental domain $W_*$ of the quantum graph $\Gamma_*$ obtained by cutting the graph $(\tilde\Gamma,\tilde A)$ of Fig.~\ref{fig:QuantumGraph3} along the central horizontal plane and retaining the upper portion.
$\Gamma_*$ consists of the square grid with vertices on $\ZZ^2$, decorated with a dangling edge (called $e'_0$ in $W_*$) attached to each vertex and parameterized by $[0,1/2]$.}
\label{fig:QuantumGraph4}}
{
\raisebox{4ex}{\scalebox{0.25}{\includegraphics{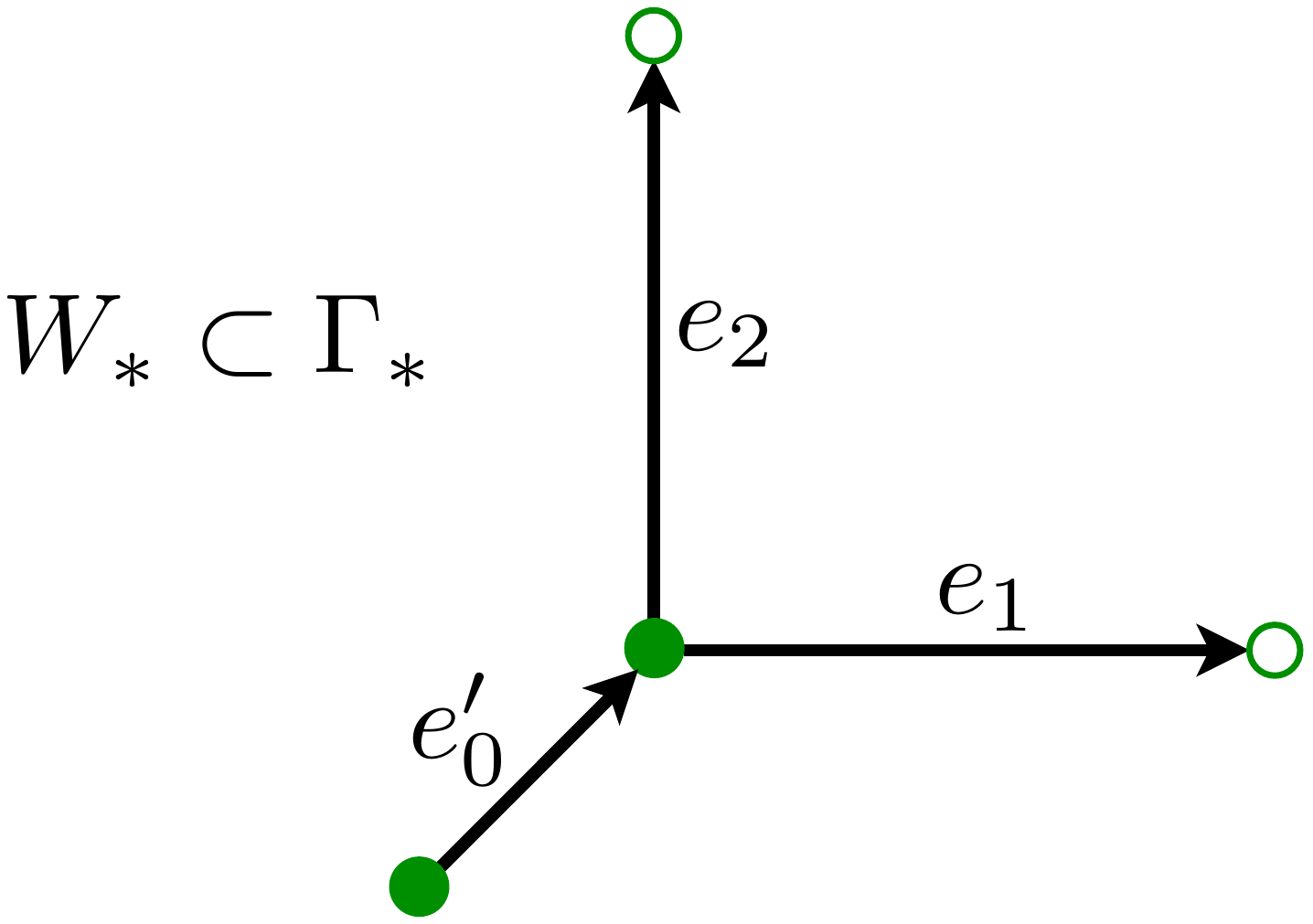}}
\hspace{1em}
}
}
\end{figure}

Let $\Gamma$ be the planar square grid whose vertex set is the integer lattice $\ZZ^2$, and let $A$ act by $-d^2/dx^2$ on each edge, with the usual continuity and zero-flux conditions.  Let $\Gamma_*$ and $A_\pm$ be defined as in section~\ref{sec:quantumcoupling}, with an edge dangling from each vertex of $\Gamma$.  The fundamental domain $W_*$ of $\Gamma_*$ is shown in Fig.~\ref{fig:QuantumGraph4}.  To determine the spectra of $A_+$ and $A_-$ on $\Gamma_*$ and the coefficients $K_{N,D}(\mu,\nu)$ defined in section~\ref{sec:quantumcoupling}, one has to solve first for the coefficients $\hat K_{N,D}=\hat K_{N,D}(\mu,\nu;k)$ by solving the following systems for $p\in W_*$ for each $z=(z_1,z_2)=(e^{ik_1},e^{ik_2})\in\TT^2$\,:
\begin{equation}
\renewcommand{\arraystretch}{1.3}
\left.
  \begin{array}{ll}
  -\hat u''_1 - \mu^2\hat u_1 = 0 & \text{on $e_1$} \\
  -\hat u''_2 - \mu^2\hat u_2 = 0 & \text{on $e_2$} \\
  -\hat u''_0 - \mu^2\hat u_0 = \renewcommand{\arraystretch}{1.1}
\left\{
  \begin{array}{ll}
    0 & \text{(homogeneous)} \\
    \cos\nu x & \text{(forced Neumann)} \\
    \frac{1}{\nu}\sin\nu x & \text{(forced Dirichlet)}
  \end{array}
\right.
& \text{on $e_0$}
  \end{array}
\right.
\end{equation}
subject to the conditions
\begin{equation}\label{matching}
\renewcommand{\arraystretch}{1.2}
\left.
  \begin{array}{l}
  \hat u_0(\onehalf) - \hat u_1(0) \,=\, 0 \\
  \hat u_0(\onehalf) - \hat u_1(1)e^{-ik_1} \,=\, 0 \\
  \hat u_0(\onehalf) - \hat u_2(0) \,=\, 0 \\
  \hat u_0(\onehalf) - \hat u_2(1)e^{-ik_2} \,=\, 0 \\
  \hat u_0'(\onehalf) - \hat u_1'(0) - \hat u_2'(0) + \hat u_1'(1)e^{-ik_1} + \hat u_2'(1)e^{-ik_2} = 0\,.    
  \end{array}
\right.
\end{equation}
The solution has the form
\begin{eqnarray*}
  \hat u_1(x) &=& \hat C_1\cos\mu x + \hat D_1{\textstyle\frac{1}{\mu}}\sin\mu x \\
  \hat u_2(x) &=& \hat C_2\cos\mu x + \hat D_2{\textstyle\frac{1}{\mu}}\sin\mu x \\
  \hat u_0(x) &=&
  \renewcommand{\arraystretch}{1.2}
\left\{
  \begin{array}{ll}
    \hat K_N\cos\mu x & \text{(Neumann homogeneous)} \\
    \vspace{-2ex}\\
    \hat K_N\cos\mu x + \displaystyle\frac{\cos\nu x}{\nu^2-\mu^2} & \text{(Neumann forced)} \\
    \vspace{-2ex}\\
    \hat K_D\frac{1}{\mu}\sin\mu x & \text{(Dirichlet homogeneous)} \\
    \vspace{-2ex}\\
    \hat K_D\frac{1}{\mu}\sin\mu x + \displaystyle\frac{\,\frac{1}{\nu}\sin\nu x\,}{\nu^2-\mu^2} & \text{(Dirichlet forced)}. \\
  \end{array}
\right.
\end{eqnarray*}

\noindent
In each of these four problems, the five conditions~(\ref{matching}) yield a system of the form
\begin{equation}\label{matrix}
  \renewcommand{\arraystretch}{1.4}
\left[
  \begin{array}{ccccc}
    a & -1 & 0 & 0 & 0 \\
    a & -\zeta_1\cos\mu & -\zeta_1\frac{1}{\mu}\sin\mu & 0 & 0 \\
    a & 0 & 0 & -1 & 0 \\
    a & 0 & 0 & -\zeta_2\cos\mu & -\zeta_2\frac{1}{\mu}\sin\mu \\
    b & -\zeta_1\mu\sin\mu & \zeta_1\cos\mu-1 & -\zeta_2\mu\sin\mu & \zeta_2\cos\mu-1
  \end{array}
\right]
\renewcommand{\arraystretch}{1.4}
\left[
  \begin{array}{c}
    \hat K \\ \hat C_1 \\ \hat D_1 \\ \hat C_2 \\ \hat D_2
  \end{array}
\right]
=
\renewcommand{\arraystretch}{1.4}
\left[
  \begin{array}{c}
    c \\ c \\ c \\ c \\ d   
  \end{array}
\right]
\frac{-1}{\nu^2-\mu^2}\,,
\end{equation}
in which the notation $\zeta_1=e^{-ik_1}$ and $\zeta_2=e^{-ik_2}$ is used for brevity.
In the forced Neumann case,
\begin{equation}
  \renewcommand{\arraystretch}{1.1}
\left.
  \begin{array}{lcl}
    a = \cos\frac{\mu}{2} && c = \cos\frac{\nu}{2} \\
    b = -\mu\sin\frac{\mu}{2} && d = -\nu\sin\frac{\nu}{2},
  \end{array}
\right.
\end{equation}
and in the forced Dirichlet case,
\begin{equation}
  \renewcommand{\arraystretch}{1.1}
\left.
  \begin{array}{lcl}
    a = \frac{1}{\mu}\sin\frac{\mu}{2} && c = \frac{1}{\nu}\sin\frac{\nu}{2} \\
    b = \cos\frac{\mu}{2} && d = \cos\frac{\nu}{2}.
  \end{array}
\right.
\end{equation}
In both unforced cases, $c=d=0$.
The determinant of the matrix in~(\ref{matrix}) is
\begin{equation}\label{determinant}
  D(a,b) = e^{-i(k_1+k_2)}{\textstyle\frac{1}{\mu}}\sin\mu
  \left[ 4a\cos\mu + b{\textstyle\frac{1}{\mu}}\sin\mu - 2a\left( \cos k_1 + \cos k_2 \right) \right].
\end{equation}
The factor $\sin\mu$ vanishes when $\lambda=\mu^2$ is a Dirichlet eigenvalue $\lambda=(\ell\pi)^2$, $\ell\in\ZZ$, of the edges $e_1$ and $e_2$.  These are exceptional eigenvalues of infinite multiplicity for both the Dirichlet and Neumann conditions at the free vertices of the graph $\Gamma_*$.

The spectrum of $(\Gamma,A)$ has no gaps---it consists of all $\lambda=\mu^2\geq0$.  This can be seen from its dispersion relation $D(1,0)=0$, or \,$\mathrm{sinc\,}\mu(4\cos\mu-2(\cos k_1 + \cos k_2))=0$.  The graph $\Gamma_*$ is obtained from $\Gamma$ by attaching a dangling edge of length $\onehalf$ as a ``decoration" to each vertex.  This causes resonant opening of gaps around the spectrum of $e'_0$ with Dirichlet condition at the vertex of attachment; see \cite{SchenkerAizenman2000} for the case of combinatorial graphs.  The gaps of $(\Gamma_*,A_-)$ are centered around the Dirichlet eigenvalues $(2\ell\pi)^2$ of $e'_0$, and the gaps of $(\Gamma_*,A_+)$ are centered around the eigenvalues $((2\ell+1)\pi)^2$ of $e'_0$ subject to endpoint conditions $u'(0)=0$ and $u(1/2)=0$, as confirmed by the calculations below.  Note that these gaps emerge within the continuous spectrum of $\Gamma$ and do not destroy the infinite-multiplicity eigenvalues $\lambda=\mu^2=(\ell\pi)^2$; they persist at the centers of the gaps in the variable $\mu$.

Excepting the values $\mu=\ell\pi$, the dispersion relation for the Neumann and Dirichlet problems are given by $D(a,b)=0$ with the appropriate values of $a$ and $b$ given above.  They boil down to
\begin{eqnarray}
  && D_{\!N}(\mu;k_1,k_2) \,:=\, 5\cos\mu - 1 - 2\left( \cos k_1 + \cos k_2 \right) = 0 
  \qquad \text{(Neumann dispersion relation)} \\  
  && D_{\!D}(\mu;k_1,k_2) \,:=\, 5\cos\mu + 1 - 2\left( \cos k_1 + \cos k_2 \right) = 0
  \qquad \text{(Dirichlet dispersion relation)}.
\end{eqnarray}
Both relations yield spectral bands and gaps.
With $\lambda=\mu^2$, and $J=[-\cos^{-1}(-3/5),\,\cos^{-1}(-3/5)]$, they are
\begin{eqnarray*}
  && \mu \in J + 2\pi\ell \hspace{1.87em} \qquad \text{bands of $A_+$ (Neumann)} \\
  && \mu \in J + \pi + 2\pi\ell \qquad \text{bands of $A_-$ (Dirichlet)}
\end{eqnarray*}
where $\ell\in\ZZ$.  Compare the result for the 1D case in Example~3.

The forced problems are solved by Cramer's rule in (\ref{matrix}) using the appropriate values of $a$, $b$, $c$, and $d$, above.  In the Dirichlet case for $\mu^2$ not in a spectral band of $A_-$, one obtains
\begin{equation*}
  \hat K_D(\mu,\nu;k_1,k_2) =
  {\frac{\mu}{(\mu^2-\nu^2)\sin\frac{\mu}{2}}}\;
  \left[
  \frac{\frac{1}{\mu}\cos\frac{\nu}{2}\sin\mu}{D_{\!D}(\mu;k_1,k_2)}
  + {\textstyle\frac{1}{\nu}\sin\frac{\nu}{2}}
  \left( 1 - \frac{1+\cos\mu}{D_{\!D}(\mu;k_1,k_2)} \right)
    \right]\,,
\end{equation*}
\begin{multline*}
  K_D(\mu,\nu) = \frac{1}{4\pi^2}\iint \hat K_D(\mu,\nu;k_1,k_2)dk_1kd_2 \\
  =\; {\frac{\mu}{(\mu^2-\nu^2)\sin\frac{\mu}{2}}}\;
  \left[
   {\textstyle\frac{1}{\mu}\cos\frac{\nu}{2}\sin\mu}\, R(\mu)
   + {\textstyle\frac{1}{\nu}\sin\frac{\nu}{2}}
   \left( 1 - (1+\cos\mu) R(\mu) \right)
  \right]\,,
\end{multline*}
in which $R(\mu)$ is a positive function defined in the spectral gaps of $A_-$, where $D_{\!D}$ is nonzero on $\TT^2$, by
\begin{equation*}
  R(\mu) \,:=\, \frac{1}{4\pi^2} \iint \frac{dk_1 dk_2}{D_{\!D}(\mu;k_1,k_2)}
    \,=\, -\frac{1}{4\pi^2} \iint \frac{dk_1 dk_2}{D_{\!N}(\mu\pm\pi;k_1,k_2)}\,.
\end{equation*}

By taking $I\subset J$ and $I\cap (J+\pi)= \emptyset$, each $\lambda$-interval $(I+2\pi\ell)^2$ is within a spectral band of $A_+$ (Neumann case) and in a spectral gap of $A_-$ (Dirichlet case).  Set $\mathring I = I\setminus\{0\}$ to exclude the exceptional eigenvalues $\lambda=(\ell\pi)^2$.

To create an anti-symmetric eigenfunction of an embedded eigenvalue of $\tilde A + \tilde V$, one simply has to create a non-embedded eigenvalue $\lambda=\mu^2$ of $A_-$ located in a spectral band of $A_+$, that is, for
$\mu\in\mathring I + 2\pi\ell$ for some $\ell\in\ZZ$.  This is possible whenever $K_{\!D}(\mu,\nu)=0$, or
\begin{equation}
  {\textstyle\nu\cot\frac{\nu}{2}} \,=\, \mu\csc\mu \left( 1+\cos\mu - R(\mu)^{-1} \right),
\end{equation}
as long as $\mu^2\not=\nu^2$.
Since the left-hand side takes on all real values, one can find $\nu$ such that $K_{\!D}(\mu,\nu)=0$ and then define the potential
\begin{equation*}
  V_0 = \mu^2-\nu^2
\end{equation*}
that realizes the bound state at $\lambda=\mu^2$ for $A_-$.

To see that the bound state decays exponentially but has unbounded support, one computes the coefficient $\hat K_D$ for all $z=(z_1,z_2)=(e^{ik_1},e^{ik_2})$,
\begin{equation*}
  \hat K_D(\mu,\nu;k_1,k_2) =
  {\frac{\mu}{(\mu^2-\nu^2)\sin\frac{\mu}{2}}}\;
  \frac{ \;4{\textstyle\frac{1}{\nu}\sin\frac{\nu}{2}}\cos\mu + {\textstyle\frac{1}{\mu}\cos\frac{\nu}{2}}\sin\mu - {\textstyle\frac{1}{\nu}\sin\frac{\nu}{2}}\left(z_1+z_1^{-1}+z_2+z_2^{-1}\right)\, }
         { 4{\textstyle\frac{1}{\mu}\sin\frac{\mu}{2}}\cos\mu + {\textstyle\frac{1}{\mu}\cos\frac{\mu}{2}}\sin\mu - {\textstyle\frac{1}{\mu}\sin\frac{\mu}{2}}\left(z_1+z_1^{-1}+z_2+z_2^{-1}\right) }\,.
\end{equation*}
For $\mu\in\mathring I + 2\pi\ell$, the denominator does not vanish on $\TT_2$, so that $\hat K_D(\mu,\nu;\cdot)$ as a function of $z=(e^{ik_1},e^{ik_2})$ is analytic in a complex neighborhood of $\TT^2$.  Similarly, the other coefficients $\hat C_{1,2}$ and $\hat D_{1,2}$ of $\hat u$ are analytic in a neighborhood of $\TT^2$.  This means that the solution $u$ itself is exponentially decaying in the lattice $\tilde\Gamma$.
But the denominator of $\hat K$ does vanish on a nonempty set in $(\CC^*)^2$, which is the Floquet surface for $A_-$.  On this set, the numerator does not vanish since $\mu\not=\nu$.  This means that $\hat K_D(\mu,\nu;\cdot)$ and therefore also $\hat u$ has singularities in $(\CC^*)^2$ so that it is not a Laurent polynomial and, hence, that $u$ does not have compact support in $\tilde\Gamma$.

\smallskip
In the other case, to create a non-embedded eigenvalue $\lambda=\mu^2$ of $A_+$ located in a spectral band of $A_-$, one needs $\mu\in\mathring I + \pi + 2\pi\ell$ for some $\ell\in\ZZ$.  Calculations yield
\begin{equation*}
  \hat K_N(\mu,\nu;k_1,k_2) =
  {\frac{1}{(\mu^2-\nu^2)\cos\frac{\mu}{2}}}\;
  \left[
  \frac{-\frac{\nu}{\mu}\sin\frac{\nu}{2}\sin\mu}{D_{\!N}(\mu;k_1,k_2)}
  + {\textstyle\cos\frac{\nu}{2}}
  \left( 1 + \frac{1-\cos\mu}{D_{\!N}(\mu;k_1,k_2)} \right)
    \right]\,,
\end{equation*}
\begin{equation*}
  K_N(\mu,\nu)
  \;=\;
  {\frac{1}{(\mu^2-\nu^2)\cos\frac{\mu}{2}}}\;
  \left[
   {\textstyle\frac{\nu}{\mu}\sin\frac{\nu}{2}\sin\mu}\, R(\mu+\pi)
   + {\textstyle\cos\frac{\nu}{2}}
   \left( 1 - (1-\cos\mu) R(\mu+\pi) \right)
  \right]\,.
\end{equation*}
Setting $K_N(\mu,\nu)=0$ yields the condition
\begin{equation*}
  {\textstyle\nu\tan\frac{\nu}{2}} \,=\, \mu\csc\mu \left( 1-\cos\mu- R(\mu+\pi)^{-1} \right)\,.
\end{equation*}
subject to $\mu^2\not=\nu^2$.

\bibliography{Shipman}

\end{document}